\documentclass[12pt, a4paper]{article}
\pdfoutput=1

\usepackage{amsmath}
\usepackage{amsfonts}
\usepackage{amssymb}
\usepackage{graphicx, rotating}
\usepackage{epsfig}
\usepackage{latexsym}
\usepackage{graphicx}
\usepackage{color}
\usepackage{amsmath,bm,amssymb}
\usepackage{cite}
\usepackage{slashed}
\usepackage{hyperref}
\hypersetup{colorlinks, citecolor=bluscuro, linkcolor=black, urlcolor=bluscuro}
\definecolor{rossos}{cmyk}{0,1,1,0.55}
\definecolor{bluscuro}{rgb}{0.15, 0.2, .85}
\definecolor{bluchiaro}{cmyk}{1,.3,0.,0.1}

\numberwithin{equation}{section}

\newcommand{\be}{\begin{equation}}
\newcommand{\ee}{\end{equation}}
\newcommand{\bea}{\begin{eqnarray}}
\newcommand{\eea}{\end{eqnarray}}

\newcommand{\Vp}{V_\pp}
\newcommand{\Ddp}{D_{d\pp}}

\def\pp{{\scriptscriptstyle +}}
\def\mm{{\scriptscriptstyle -}}

\newcommand{\arXiv}[2]{\href{http://arxiv.org/pdf/#1}{{\tt [#2/#1]}}}
\newcommand{\arXivold}[1]{\href{http://arxiv.org/pdf/hep-#1}{{\tt [#1]}}}

\begin{document}
\allowdisplaybreaks
\begin{titlepage}
\begin{flushright}
IFT-UAM/CSIC-22-144
\end{flushright}
\vspace{.3in}

\vspace{1cm}
\begin{center}
{\Large\bf\color{black} 
Vacuum Decay Actions from Tunneling Potentials\vspace*{0.15cm}\\ for General Spacetime Dimension
} \\
\vspace{1cm}{
{\large J.R.~Espinosa$^1$} and {\large J.-F. Fortin$^2$}
} \\[7mm]
{\it ${}^1$ Instituto de F\'{\i}sica Te\'orica IFT-UAM/CSIC, \\ 
C/ Nicol\'as Cabrera 13-15, Campus de Cantoblanco, 28049, Madrid, Spain\\
${}^2$ D\'epartement de Physique, de G\'enie Physique et d'Optique\\
Universit\'e Laval, Qu\'ebec, QC G1V 0A6, Canada
}

\end{center}
\bigskip

\vspace{.4cm}

\begin{abstract}
The tunneling potential method to calculate the action for vacuum decay is an alternative to the Euclidean bounce method that has a number of attractive features. In this paper we extend the formalism to general spacetime dimension $d>2$ and use it to give simple proofs of several results. For Minkowski or Anti de Sitter false vacua, we show that gravity or higher barriers increase vacuum lifetime and describe a very clean picture of gravitational 
quenching of vacuum decay. We also derive the thin-wall limit of the action, show how detailed balance for dS to dS transitions works in the new formalism and how to obtain potentials for which the vacuum decay solution can be obtained analytically.

\end{abstract}
\bigskip

\end{titlepage}

\section{Introduction \label{sec:intro}}

Long-lived false vacua appear often in particle physics models (from the Standard Model to the string theory landscape) with decay rates per unit volume suppressed by the exponential of (minus) a tunneling action. Traditionally, this action is computed by the elegant and powerful Euclidean method (developed by Coleman and collaborators \cite{Coleman,CdL}) in terms of a bounce configuration that lives in Euclidean spacetime. 

A new approach to the calculation of such tunneling actions which does not rely on Euclidean bounces has been developed in \cite{E}.
This so-called ``tunneling potential approach'' reformulates the tunneling action calculation as a simple variational problem in field space. Instead of a bounce one has to find a ``tunneling potential'' function, $V_t(\phi)$, that interpolates between the false vacuum and (the basin of) the true vacuum and minimizes an action functional, an integral in field space of a simple action density. The resulting action reproduces the Euclidean result and the approach has a number of good properties: it allows a fast and precise numerical calculation of the action \cite{E}; it can be adapted to the study of vacuum decay by thermal fluctuations \cite{E}; it can be used to get solvable potentials (that permit the analytical solution of the  tunneling problem) \cite{E,Eg,EFH}; it is very useful for vacuum decay in multi-field potentials as one searches for a minimum of the action functional (instead of a saddle-point, as in the Euclidean method) \cite{EK}; it can be generalized quite simply to include gravitational corrections offering a quite direct route to the derivation of key results \cite{Eg,Estab}; it can deal with issues of gauge invariance \cite{gauge}; etc.

Previous work on the tunneling potential formalism has been mostly done for $d=4$ spacetime dimensions, although the tunneling action for general spacetime dimension $d$ was derived in \cite{E} for the case without gravitational corrections. The purpose of this paper is to derive the tunneling action for general $d$ including gravity and to discuss some results. The generalization is straightforward as there are no qualitative changes in behavior but rather a quantitative dependence on $d$. We restrict the discussion to $d>2$ as the $d=2$ case requires special modifications (to be nontrivial) and deserves a separate study (see \cite{PQ} and references therein for a recent discussion of vacuum transitions in $d=2$). We also leave aside spacetimes with compactified dimensions, in which new instabilities like bubbles of nothing \cite{WBoN} can trigger spacetime decay. Interestingly, tunneling potentials can be applied successfully to the study of such instantons \cite{BEHS}.

The structure of the paper is as follows. In section \ref{sec:action} we present our main result, the tunneling action density in $V_t$ formalism for general $d$ and including gravity, equation (\ref{sd}). We obtained this action by reverse engineering the differential equation for $V_t$ but it can also be obtained using a canonical transformation that relates Euclidean and $V_t$ formulations (this is done in Appendix~\ref{app:cantransf}). Although the action (\ref{sd}) is simply expressed in terms of a hypergeometric function, it is possible to rewrite it in terms of elementary functions. This is done in \ref{sec:action} for $d=3,4,5$
and in general in Appendix~\ref{app:elementary}. 

Although the equivalence between the Euclidean and $V_t$ formulations follows from the canonical transformation that relates them, it is illustrative to show in detail how this equivalence comes about, and this is done in  Appendix~\ref{app:equiv}. While the action  (\ref{sd}) assumes a canonical kinetic term for a  scalar minimally coupled to gravity, both assumptions can be lifted, and this is done in Appendix~\ref{app:gen}.

We then use the new action to prove a number of results in a very simple manner. First, we prove in Section~\ref{sec:results} that  Anti de Sitter (AdS) or Minkowski false vacua are made more stable by higher barriers and by gravitational effects. If the latter are strong enough, they can eventually stabilize the false vacua completely (gravitational quenching of the decay). The $V_t$ formalism is particularly well suited to describe this quenching effect. Second, for transitions between de Sitter (dS) vacua
there is a simple (detailed balance) relation for the difference between back and forth transitions and, in Section~\ref{sec:detbal} we show how this relation arises in the $V_t$ formalism.

Section~\ref{sec:thw} derives the thin-wall tunneling action in $V_t$ formalism (to be compared with the derivation in the Euclidean approach, which is relegated to Appendix~\ref{app:thwE}) illustrating how the thin-wall regime behaves for AdS or Minkowski vacua compared to the dS case. Finally, Section~\ref{sec:exact} explains how to use the $V_t$ formalism to generate analytically solvable examples of vacuum decay for general $d$, illustrating the technique with a simple example.

\section{Tunneling Action for General Dimension  \label{sec:action}}

In the formulation due to Coleman and De Luccia \cite{CdL}, false  vacuum decay is
described by a bounce configuration that extremizes the Euclidean action, which in $d$ dimensions reads
\be
S_E=\int d^dx\sqrt{g}\left[-\frac{1}{2\kappa}R+\frac12g^{\mu\nu}\partial_\mu\phi\partial_\nu\phi+V(\phi)\right]+S_{\rm GHY}\ ,
\label{SE0}
\ee
where $S_{GHY}$ is the Gibbons-Hawking-York boundary action \cite{GH,Y}. Here $\kappa=1/m_p^{d-2}$, with $m_p$ the reduced Planck mass.

We assume that the Coleman-De Luccia (CdL) bounce in an Euclidean space of $d$ dimensions has $O(d)$ symmetry, so that the scalar field depends
on a single radial variable, $\phi(\xi)$, while the metric can be written as
\be
ds^2 = d\xi^2+\rho(\xi)^2d\Omega^2_{d-1}\ ,
\ee 
and depends on a single function $\rho$. Here $\xi$ is the  coordinate from the center of the bounce, measuring the radial distance along lines normal to $d-1$ spheres of radius of curvature $\rho(\xi)$ and $d\Omega^2_{d-1}$ is the line element for a unit $d-1$ sphere. With this assumption, one has
\be
R=-2(d-1)\frac{\ddot \rho}{\rho}+(d-1)(d-2)\frac{(1-\dot\rho^2)}{\rho^2}\ ,
\label{R}
\ee
where $\dot x\equiv dx/d\xi$. The $\ddot\rho$ term is cancelled, after integrating by parts, by $S_{\rm GHY}$, which is
\be
S_{\rm GHY}=-\left.\frac{(d-1)}{\kappa}V_{S,d-1}\rho^{d-2}
\dot\rho\,\right|_{\xi=0}^{\xi=\xi_e}\ ,
\ee
and the action takes the form
\be
S_E[\phi,\rho]=V_{S,d-1}\int_0^{\xi_e}
\left[\rho^{d-1}\left(\frac12\dot\phi^2+V\right)-\frac{(d-1)(d-2)}{2\kappa}\rho^{d-3}(1+\dot\rho^2)\right]d\xi\ ,
\label{SE}
\ee
where 
$\xi_e$ is $\infty$ for AdS or Minkowski and finite for dS and 
\be
V_{S,d-1}=\frac{2\pi^{d/2}}{\Gamma(d/2)}\ ,
\ee
is the volume of a $d-1$ sphere of unit radius.

The CdL equations, derived as Euler-Lagrange equations from the extremality of the Euclidean action,  are
\bea
&&\ddot\phi+(d-1)\frac{\dot\rho}{\rho}\dot\phi=V'\ ,\\
\label{CdL1}
&&\dot\rho^2=1+\frac{2\kappa\rho^2}{(d-1)(d-2)}\left(\frac12\dot\phi^2-V\right)\ ,
\label{CdL2}
\eea
where $x'\equiv dx/d\phi$. 

As is well known, the bounce for the decay of a vacuum at $\phi_\pp$ satisfies the boundary conditions
\bea
&&
\phi(0)=\phi_0\ , \quad \phi(\infty)=\phi_\pp\ , \quad
\dot\phi(0)=\dot\phi(\infty)=0\ ,\nonumber\\
&&
\rho(0)=0\ , \quad \rho(\infty)=\infty\ ,\quad \dot\rho(0)=1\ ,
\eea
for an AdS or Minkowski vacuum, while for a dS vacuum we have instead
\bea
&&
\phi(0)=\phi_0\ , \quad \phi(\xi_e)=\phi_{0,\pp}\neq \phi_\pp\ , \quad
\dot\phi(0)=\dot\phi(\xi_e)=0\ ,\nonumber\\
&&
\rho(0)=\rho(\xi_e)=0\ ,\quad \dot\rho(0)=1\ ,\quad \dot\rho(\xi_e)=-1\ .\label{BCdS}
\eea
In the equations above,  $\phi_0$ is a field value to be found so as to satisfy the other boundary conditions. 

The Euclidean tunneling action is the difference between the action of the CdL instanton and the background false vacuum action
\be
\Delta S_E = S_E[\phi,\rho] - S_E[\phi_\pp,\rho_\pp]\ ,
\ee
where $\phi_\pp$ and $\rho_\pp$ are, respectively, the field value and metric at the false vacuum.

In the tunneling potential formulation, we get rid of all Euclidean quantities and describe the tunneling configuration in terms of a single function $V_t(\phi)$, the tunneling potential, which is related to Euclidean quantities by \cite{E}
\be
V_t = V -\frac12 \dot\phi^2\ ,
\label{Vt}
\ee
where $\dot\phi$ is evaluated with the CdL profile and then expressed as a function of $\phi$. Using the CdL equations, the metric function $\rho(\xi)$ can be related to $V_t$ quantities  as 
\be
\rho = (d-1) \frac{\sqrt{2(V-V_t)}}{D_d}\ ,
\label{rho}
\ee
with
\be
D_d \equiv \sqrt{V_t'{}^2+4\kappa_d (V-V_t)V_t}\ ,
\label{Dd}
\ee
and we have defined 
\be
\kappa_d\equiv \kappa\frac{(d-1)}{(d-2)}\ .
\ee
We also have
\be
\dot\rho =-\frac{V_t'}{D_d}\ .
\label{drho}
\ee

Although we have taken the Euclidean action as starting point of our derivation and introduced $V_t$ relating it to the Euclidean bounce, this is just a convenient shortcut to derive the $V_t$ formulation, in which $V_t$ is the unknown function to be obtained, instead of the Euclidean bounce.
The differential ``equation of motion'' (EoM) for $V_t$ follows from the $\xi$ derivative of (\ref{rho}) and reads:
\be
2(V-V_t)\left\{V_t''+2\kappa \left[\frac{(d-1)}{(d-2)}V-V_t\right]\right\}+\left(\frac{d}{d-1}V_t'-V' \right)V_t'=0\ .
\label{EoMVt}
\ee
Alternatively, this EoM can be rewritten in a simpler form in terms of $D_d$ as
\be
\frac{d}{d\phi}\log D_d = \frac{1}{2(V-V_t)}\left(V'-\frac{d\, V_t' }{d-1} \right)\ .
\label{EoMVtD}
\ee

The boundary conditions for $V_t$ are
\be
V_t(\phi_\pp)=V(\phi_\pp)\ ,\quad 
V_t(\phi_0)=V(\phi_0)\ ,\quad
V'_t(\phi_\pp)=0\ ,\quad
V'_t(\phi_0)=\frac34 V'(\phi_0)\ .
\ee
For AdS or Minkowski false vacua (or in the absence of gravity) 
the function $V_t(\phi)$ monotonically decreases away from $\phi_\pp$. For dS, instead, it first grows and later decreases \cite{Eg}. In fact, in the interval from $\phi_\pp$ to $\phi_{0\pp}$, one has $V_t=V$ \cite{Eg}.

For later use it is convenient to define the quantities
\be
R(\phi)^2\equiv\frac{(d-1)(d-2)}{2\kappa |V(\phi)|}\ ,\quad
R_t(\phi)^2\equiv\frac{(d-1)(d-2)}{2\kappa |V_t(\phi)|}\ .
\ee
At potential minima, $R(\phi)$ gives the corresponding dS or AdS radius (divergent in Minkowski).

The action density whose variation gives the EoM above can be obtained  (up to a multiplicative constant and a boundary term we fix so as to agree with the tunneling Euclidean action, $\Delta S_E$) using standard techniques.\footnote{Alternatively, one can use a canonical transformation that relates $V_t$ and Euclidean quantities to get the new action \cite{EJK} (see Appendix~\ref{app:cantransf}). Both methods give the same answer.} We get
\be
{\it s}^{(d)}=\frac{\pi^{(d+1)/2}R_t^d}{\Gamma[(d+1)/2]}
(V_t'+|V_t'|)
+
\, _2F_1\left(\frac{d-1}{2},\frac{d}{2};\frac{d}{2}+1;1-\frac{D_d^2}{V_t'{}^2}\right) \, {\it s}_0^{(d)}
\label{sd}
\ee
with
\be
{\it s}_0^{(d)}=
\frac{(d-1)^{d-1}\left[2\pi(V-V_t)\right]^{d/2}}{\Gamma(1+d/2)|V_t'|^{d-1}}\ .
\label{s0d}
\ee
Some comments on this result are in order.
We see that $s^{(d)}$ is positive definite.
In the limit $\kappa\rightarrow 0$ (no gravity) one simply gets 
${\it s}^{(d)}={\it s}_0^{(d)}$, (as $V_t'\leq 0$ and $1-D^2/V_t'{}^2\rightarrow 0 $). In estimating the gravitational impact on tunneling, however, one should keep in mind that $V_t$ is different
with and without gravity. 

The hypergeometric function in (\ref{sd}) can be expressed in terms of elementary functions given that $d$ is an integer.
For $d=4$ one recovers the result \cite{Eg}
\be
{\it s}^{(4)} =\frac{6\pi^2}{\kappa^2}
\frac{(D_4+V_t')^2}{V_t^2 D_4}\ ,
\ee
with
$D_4^2=V_t'{}^2+6\kappa(V-V_t)V_t$.

For $d=3$ one gets
\be
{\it s}^{(3)} =\frac{\pi}{\kappa^{3/2}}\left\{\pi\frac{(V_t'+
|V_t'|)}{|V_t|^{3/2}}+\frac{4}{V_t}\sqrt{2\kappa(V-V_t)}\left[1-\frac{{\rm Arctanh}(\sqrt{1-D_3^2/V_t'{}^2})}{\sqrt{1-D_3^2/V_t'{}^2}}\right]\right\}\ ,
\ee
with 
$D_3^2=V_t'{}^2+8\kappa(V-V_t)V_t$.

For $d=5$ one gets
\be
{\it s}^{(5)} =\frac{\pi^2}{\kappa^{5/2}}\left\{\frac{\pi}{2}\frac{(V_t'+
|V_t'|)}{(|V_t|/6)^{5/2}}+\frac{48\sqrt{2\kappa(V-V_t)}}{V_t^2}\left[2+\frac{V_t'{}^2}{D_5^2}-\frac{3{\rm Arctanh}(\sqrt{1-D_5^2/V_t'{}^2})}{\sqrt{1-D_5^2/V_t'{}^2}}\right]\right\}\ ,
\ee
with 
$D_5^2=V_t'{}^2+16\kappa(V-V_t)V_t/3$. Appendix \ref{app:elementary} explains how to obtain a general expression for $s^{(d)}$ in terms of elementary functions for arbitrary $d>2$.
 
The final expression for the tunneling action is 
\be
S=\int_{\phi_\pp}^{\phi_0}s^{(d)}(\phi)d\phi\ .
\ee
Here $\phi_\pp$ (which can be taken to be 0 without loss of generality) is the false vacuum field value and $\phi_0$ has to be found minimizing the integral [it corresponds to $\phi(0)$ of the Euclidean formalism]. In our convention, $\phi_\pp<\phi_0<\phi_\mm$, with $\phi_\mm$ being the true vacuum.
For Minkowski or AdS false vacua one has $V_t'\leq 0$ so that the first term in (\ref{sd}) vanishes and one gets
\be
S_{\rm Mink,AdS}=\frac{(d-1)^{d-1}\left[2\pi\right]^{d/2}}{\Gamma(1+d/2)}\int_{\phi_\pp}^{\phi_0} \frac{(V-V_t)^{d/2}}{|V_t'|^{d-1}}
{}_2F_1\left(\frac{d-1}{2},\frac{d}{2};\frac{d+2}{2};1-\frac{D_d^2}{V_t'{}^2}\right) 
d\phi\ .
\label{SminkAdS}
\ee
For dS false vacua there is a region in field space, between $\phi=\phi_\pp$ and some $\phi_{0\pp}<\phi_0$ where $V_t(\phi)=V(\phi)$, while the interval $(\phi_{0\pp},\phi_0)$ corresponds to the CdL field range and $V_t<V$ is non trivial. The action density in the interval $(\phi_\pp,\phi_{0\pp})$ takes the simple form
\be
{\it s}^{(d)}=\frac{2\sqrt{\pi}V_t'}{\Gamma[(d+1)/2]}\left[
\frac{(d-1)(d-2)\pi}{2\kappa V_t}\right]^{d/2}
\ ,\quad\quad (\mathrm{for}\ \phi_\pp\leq\phi\leq\phi_{0\pp})\ ,
\label{sdtrivial}
\ee
and can be integrated exactly. Then the tunneling action for dS false vacua reads
\be
S_{\rm dS}=\frac{4\pi^{(d+1)/2}}{\kappa\Gamma[(d-1)/2]}
\left(R_\pp^{d-2}-R_{0\pp}^{d-2}\right)
+\int_{\phi_{0\pp}}^{\phi_0}s^{(d)}(\phi)d\phi\ ,
\label{SdS}
\ee
with $V_\pp\equiv V(\phi_\pp)$ and $V_{0\pp}\equiv V(\phi_{0\pp})$ and $s^{(d)}$ as given in (\ref{sd}).
An alternative expression for this action is 
\be
S_{\rm dS}=\frac{4\pi^{(d+1)/2}}{\kappa\Gamma[(d-1)/2]}
\left(R_\pp^{d-2}-R_T^{d-2}\right)
+\int_{\phi_{0\pp}}^{\phi_0}s^{(d)}_{CdL}(\phi)d\phi\ ,
\label{SdSalt}
\ee
where $s^{(d)}_{CdL}(\phi)$ is just the part of (\ref{sd}) that depends on the hypergeometric function, and $R_T\equiv R_t(\phi_T)$,
where $\phi_T$ is the field value at which $V_t$ reaches its maximum, so that $V_t'(\phi_T)=0$.

In appendix \ref{app:equiv} we show the equivalence between the tunneling actions calculated in Euclidean and tunneling potential formalisms.
In appendix \ref{app:gen} we extend the result of this section to a more general action in which we allow for a field-dependent noncanonical kinetic term and a nonminimal coupling of the scalar field to gravity.

\section{Some General Results for Minkowski or AdS Vacua\label{sec:results}}

From the general expression of the tunneling action density in (\ref{sd}) one can generalize to arbitrary dimension ($d>2$) results known for the $d=4$ case. In the context of the tunneling potential approach, these results were proven  in \cite{Estab}.
 The results for the decay of Minkowski or AdS vacua are discussed in the following subsections. 
 
\subsection{Higher Barriers Make False Vacua More Stable}
Consider two potentials that take the same values at a false (Minkowski or AdS) minimum\footnote{We discuss here the case of vacua which are local minima. The case of decaying AdS maxima (that can be stable if they respect the Breitenlohner-Freedman bound \cite{BF}) is more subtle, see {\it e.g.}  \cite{FHR}.} located at $\phi_\pp$ and a true AdS vacuum at $\phi_\mm$ but with $V_2\geq V_1$ in between. Take any $V_t$ that leads to a finite tunneling action for the decay in potential $V_2$. The corresponding tunneling action densities satisfy the inequality
\be
{\it s}_{2}(V_t) \geq {\it s}_{1}(V_t)
\ ,\label{s2geqs1}
\ee
with
\be
{\it s}_{i}(V_t)\equiv \frac{(d-1)^{d-1}}{\Gamma(1+d/2)}\left[\frac{\pi V'_t{}^2}{2\kappa_d(-V_t)}\right]^{d/2}F_d(x_i)\geq 0\ ,
\ee
where
\be
F_d(x_i)\equiv (1-x_i^2)^{d/2}\, {}_2 F_1\left(\frac{d-1}{2},\frac{d}{2};\frac{d+2}{2};1-x_i^2\right)\ ,
\ee
contains all the dependence on $V_i$ via
\be
x_i^2\equiv \frac{D_{d,i}^2}{V_t'{}^2}\equiv \frac{1}{V_t'{}^2}\left[V_t'^2+4\kappa_d(V_i-V_t)V_t\right]\ .
\ee
As $V_t\leq 0$ and $V_2\geq V_1$ we have $0\leq x_i\leq 1$ and $x_2\leq x_1$. The inequality (\ref{s2geqs1}) then follows from the fact that $F_d(x)$ is a monotonically decreasing function in $(0,1)$,
as proven by
\be
\frac{dF_d(x)}{dx}=-d \left(\frac{1}{x^2}-1\right)^{(d-2)/2}\leq 0\ .
\ee

After having established (\ref{s2geqs1}), the proof follows the same logic of the $d=4$ case \cite{Estab}. Let $V_{t\kappa,i}$ be the tunneling potentials for the $V_i$ ({\it i.e.} $V_{t\kappa,i}$ give the minimum of the respective actions). The $V_{t\kappa,i}$ are defined in some intervals $(\phi_\pp,\phi_{0,i})$, where $\phi_{0,i}\leq \phi_\mm$ are the exit points of the tunneling. Then, $V_{t\kappa,2}$ intersects the lower potential $V_1$ at some field value $\phi_{0,21}\leq \phi_{0,2}$ and we have
\bea
S_{2}[V_{t\kappa 2}]&\equiv &
\int_{\phi_\pp}^{\phi_{0,2}}{\it s}_{2}(V_{t\kappa 2})d\phi\geq 
\int_{\phi_\pp}^{\phi_{0,21}}{\it s}_{2}(V_{t\kappa 2})d\phi
\nonumber\\
&\geq & 
\int_{\phi_\pp}^{\phi_{0,21}}{\it s}_{ 1}(V_{t\kappa 2})d\phi
\geq 
\int_{\phi_\pp}^{\phi_{0,1}}{\it s}_{ 1}(V_{t\kappa 1})d\phi\equiv S_{1}[V_{t\kappa,1}] \ .
\eea
The first inequality follows from $\phi_{0,2}\geq\phi_{0,21}$ plus the positivity of the action density; the second inequality follows from (\ref{s2geqs1}); and the third from the fact that $V_{t\kappa 1}$ minimizes the action for $V_1$.  Notice that the upper limits in the last two integrals correspond to the points where $V_{t\kappa 2}(\phi_{0,21})=V_1(\phi_{0,21})$ and $V_{t\kappa 1}(\phi_{0,1})=V_1(\phi_{0,1})$. As was the case for $d=4$,
the argument does not require the inequality to hold for the action densities, ${\it s}_2(V_{t\kappa 2})\geq {\it s}_1(V_{t\kappa 1})$, which can be violated.

\subsection{Gravity Makes Vacua More Stable}
To prove that gravity makes false (Minkowski or AdS) vacua more stable, take a path $V_t(\phi)$ out of the metastable vacuum 
$\phi_\pp$ of a  potential $V(\phi)$. The corresponding tunneling action densities with and without gravity satisfy 
\be
{\it s}(V_t) \geq 
{\it s}_0(V_t)\ .\label{sgeqs0}
\ee
This inequality follows immediately from (\ref{sd}), that gives  the simple relation
\be
\frac{s(V_t)}{{\it s}_0(V_t)}=
{}_2F_1\left(\frac{d-1}{2},\frac{d}{2};\frac{d+2}{2};1-x^2\right)\geq 1,
\ee
with $x^2\equiv D_d^2/V_t'{}^2$ satisfying\footnote{One has $V_t,V_t'\leq 0$ and $D_d$ must be real for $V_t$ to be a  decay path allowed by gravity, see next subsection.} $0\leq x^2\leq 1$.
The hypergeometric function is bigger than 1 as it is defined by a convergent series of the form  $1+\sum_{n=1}^\infty a_n z^n$ with $a_n>0$.

To finish the proof, let $V_{t\kappa}$ and $V_{t0}$ be the tunneling potentials that minimize the actions with and without gravity, respectively. Then 
\be
S[V_{t\kappa}] \equiv \int_{\phi_\pp}^{\phi_0}{\it s}(V_{t\kappa}) d\phi\geq \int_{\phi_\pp}^{\phi_0}{\it s}_0(V_{t\kappa})d\phi\equiv S_0[V_{t\kappa}]\geq S_0[V_{t0}]\ ,
\ee
where the first inequality follows from (\ref{sgeqs0}) and the second from the fact that the tunneling functional $S_0[V_t]$ is minimized by $V_{t0}$.

\subsection{Gravitational Quenching\label{GQ}}

One of the most striking effects of gravity on vacuum decay is
gravitational quenching: the fact that gravitational effects can  stabilize completely false vacua and lead to an infinite tunneling action that forbids vacuum decay \cite{CdL}. The effect happens for any spacetime dimension $d>2$ and we show below how this is described by the $V_t$ formalism.

In the $V_t$ formulation an allowed vacuum decay  needs to satisfy the condition 
\be
D_d^2 = V_t'{}^2+4\kappa_d(V-V_t)V_t>0\ ,
\label{quench}
\ee
with gravitational quenching of the decay happening when this condition cannot be satisfied no matter how $V_t$ is chosen \cite{Eg}. The impact of gravitational effects is increased by making $\kappa$ larger (in fact, what is increased is the dimensionless combination $\kappa M^2$, where $M$ is a characteristic mass scale of the potential). For Minkowski or AdS vacua the second term in (\ref{quench}) is negative and for large enough $\kappa$ it will be impossible to satisfy the condition
(\ref{quench}) for any $V_t$: the potential will be stabilized. 
Compared with the $d=4$ case, it is clear that general $d$ does not introduce a qualitative difference: it simply modifies
 slightly the impact of gravitational corrections, with the factor $4\kappa_d $ decreasing from $8\kappa$ for $d=3$ to $\simeq 4\kappa$ for $d\gg 1$.

The expected parametric behaviour needed for quenched potentials is the same as it was for $d=4$ \cite{Estab}: large $\Delta\phi\equiv \phi_\mm-\phi_\pp$, high potential barriers, shallow true minima or deep AdS false minima. In more quantitative detail, one can interpret the condition $D_d^2>0$ needed for AdS or Minkowski vacuum decay with gravity as implying that $V_t$ must satisfy a condition stronger than monotonicity to have $D_d$ real:
\be
V_t'\leq -\sqrt{4\kappa_d(V-V_t)(-V_t)}\ .
\label{newmono}
\ee

For a potential $V$ with a metastable minimum at $\phi_\pp$ we can get a ``critical'' tunneling potential, $V_{tc}$, as the solution to  $D_d\equiv 0$, that is
\be
V_{tc}'= -\sqrt{4\kappa_d(V-V_{tc})(-V_{tc})}\ ,
\label{Vtc}
\ee
with boundary condition $V_{tc}(\phi_\pp)=V(\phi_\pp)\equiv V_\pp$.
To integrate (\ref{Vtc}) and get $V_{tc}$, it is enough to have the boundary condition at $\phi_\pp$. Other solutions of (\ref{Vtc}) for different boundary values of $V_{tc}(\phi_\pp)$ generate a family of non-intersecting integral curves for $D=0$ that cover 
the area below $\mathrm{Min}\{V_\pp,V\}$. 

Depending on the strength of gravitational effects one can distinguish three different cases, exactly as in $d=4$:

{\bf Subcritical case.} This is the typical case with weak gravitational effects: the critical $V_{tc}$ deviates a bit from being horizontal but reaches $V$ at some field value well below $\phi_\mm$ and $V_t$ can lie below $V_{tc}$, intersect the $D=0$ integral lines from above [so as to satisfy (\ref{newmono})] and hit $V$ at $\phi_0$ with finite action. In these subcritical case, gravity makes the false $\phi_\pp$ vacuum more stable without forbidding its decay.  

{\bf Critical case.} In this special case, $V_t\equiv V_{tc}$. Note that $D_d=0$ gives a solution to the EoM  (\ref{EoMVtD}) so that $V_{tc}$ satisfies the right boundary condition at $\phi_0=\phi_\mm$. For this critical case, the tunneling action is infinite as the hypergeometric function
in (\ref{sd}) diverges for $D_d\rightarrow 0$ as [with $x=D_d/(-V_t')$]
\be
{}_2F_1\left(\frac{d-1}{2},\frac{d}{2};\frac{d+2}{2};1-x^2\right)=
\begin{cases}
-3[1+\ln(x/2)]+{\cal O}(x^2)\ , &  \mathrm{for}\quad d=3\ ,\\
\displaystyle{\frac{d}{(d-3)x^{d-3}}} 
+{\cal O}(x^{5-d})+{\cal O}(x^0)\ , & \mathrm{for}\quad d>3\ ,
\end{cases}
\ee
so that gravity forbids the decay of $\phi_\pp$ into $\phi_\mm$. This critical case corresponds to the so-called ``great divide'' case of \cite{GDivide} (discussed in that paper for $d=4$, see also \cite{BFL}). 

Between two vacua connected by $V_{tc}$ one can have a static domain wall which is the infinite radius limit of a Coleman-De Luccia (CdL) bubble. The domain wall tension can be obtained as
\be
\sigma\equiv\int_{\phi_\pp}^{\phi_\mm}\sqrt{2(V-V_t)}d\phi=
\int_{\phi_\pp}^{\phi_\mm}
\frac{-V_t'}{\sqrt{-2\kappa_dV_t}}
d\phi=\left.\sqrt{\frac{-2V(\phi)}{\kappa_d}}
\right|^{\phi_\mm}_{\phi_\pp}\ ,
\label{wtdw}
\ee
where we have used $D_d=0$ (with $V_t'\leq 0$) to write the second expression and $V_t(\phi_\pm)=V(\phi_\pm)$ to write the last. The domain-wall field profile $\phi_{DW}$ can be obtained from $V_{tc}$ inverting $V_{tc}=V-\dot\phi_{DW}^2/2$, see \cite{Estab}.

One can also solve $D_d\equiv 0$ for $V$ to obtain that any potential made critical by gravity takes the generic form
\be
V_c(\phi)=V_t-\frac{V_t'{}^2}{4\kappa_d V_t}\ ,
\label{Vc}
\ee
for a monotonic function $V_t(\phi)$. 
This formula reproduces in a straightforward way the old results of \cite{Boucher,AP,T}. Supersymmetric potentials are naturally of this critical form.

In the context of cobordism, the condition $D_d=0$ corresponds to end-of-the world branes, and has been studied, using the tunneling potential formalism, in \cite{ACDHU}.

{\bf Supercritical case.} For even stronger gravitational effects the $V_{tc}$ potential is curved down so much that it does not intersect $V$ after leaving from $\phi_\pp$. As $V_t$ should lie below $V_{tc}$, this prevents the existence of a viable $V_t$ with real $D_d$ and vacuum decay is again forbidden by gravity \cite{CdL,quench2}.

To sum up, in order to find out if a given (Minkowski or AdS) false vacuum can decay one solves (\ref{Vtc}) with $V_{tc}(\phi_\pp)=V_\pp$ and checks if $V_{tc}$ intersects $V$ or not. If it does, decay is allowed; if it does not, decay is quenched. The critical case $V_t=V_{tc}$
corresponds to an intersection precisely at the minimum $\phi_\mm$ with vacuum decay forbidden and $V_{tc}$ describing a domain-wall between the two vacua.

Finally, from the $d$-dependence of $D_d$  [so that $(d-1)/(d-2)$ is a monotonically decreasing function of $d$] we can also see that a potential that is critical at dimension $d$ is subcritical at dimension $d+1$ (as the critical $V_{tc}$ for dimension $d$ has now finite action) and supercritical at $d-1$ (as the critical $V_{tc}$ should lie below the one for dimension $d$ and therefore it no longer intersects $V$). See section \ref{sec:thw} for an illustration of this point.

\section{Detailed Balance for dS to dS Transitions \label{sec:detbal}}

In the decay of a dS vacuum, only the finite space inside the horizon is required to transition. This makes the rate non zero generically and allows upwards transitions, from a dS vacuum to another with higher cosmological constant. For the discussion below, it is convenient to rewrite the tunneling action $S_{\pp\mm}$ for the decay from a dS vacuum at $\phi_\pp$ to a dS vacuum at $\phi_\mm$ as the integral of the action density 
(\ref{sd}) in the full interval from $\phi_\pp$ to $\phi_\mm$. The action has three different pieces \cite{Eg}: In the first, from $\phi_\pp$ to some $\phi_{0\pp}$, one has $V_t\equiv V$, with $V_t'>0$. This gives $D_d^2=V_t'{}^2$ and $s_0^{(d)}=0$, so that  the action density is simply as given in (\ref{sdtrivial}).
From $\phi_{0\pp}$ to some $\phi_{0\mm}$ one has $V_t<V$ and this range corresponds to the field range of the CdL Euclidean bounce. Finally, from $\phi_{0\mm}$ to $\phi_{\mm}$ one can take again $V_t\equiv V$, with $V_t'<0$, which simplifies the tunneling action density to ${\it s}^{(d)}=0$.

The decay in the opposite direction, from $\phi_\mm$ to $\phi_\pp$, with action $S_{\mm\pp}$,  proceeds in a similar manner, in fact with the same $V_t$ function, but now taken as starting from $\phi_\mm$, so that its derivative flips sign. This implies that there is in $S_{\mm\pp}$ a simple non-zero contribution from the interval $\phi_\mm$  to $\phi_{0\mm}$ and a zero contribution from the interval from $\phi_{0\pp}$ to $\phi_\pp$.

The difference between the two tunneling actions, $\Delta S\equiv S_{\pp\mm}-S_{\mm\pp}$, takes a very simple form, as only the term linear in $V_t'$ in (\ref{sd}), the only one that flips sign, contributes. This term can be integrated exactly and one gets
\be
\Delta S =2\int_{\phi_\pp}^{\phi_\mm}\frac{\sqrt{\pi}}{\Gamma(\frac{d+1}{2})}\left[\frac{(d-1)(d-2)\pi}{2\kappa V_t}\right]^{d/2}V_t'\, d\phi =\left.\frac{-4\sqrt{\pi}\, V}{(d-2)\Gamma(\frac{d+1}{2})}\left[\frac{\pi(d-1)(d-2)}{2\kappa V}\right]^{d/2}\right|_{\phi_\pp}^{\phi_\mm}\ .
\ee 
This can be rewritten simply as
\be
\Delta S = {\cal S}_\pp - {\cal S}_\mm\ ,
\ee
where ${\cal S}_\pm$ is the Gibbons-Hawking entropy of a dS vacuum with cosmological constant $V_\pm$ \cite{GH}. Indeed, this entropy is one fourth of the horizon's area in Planck units
\be
{\cal S}_\pm=\frac14 \frac{A_\pm}{l_P^{d-2}}\ ,
\ee
where the area is given by $A_\pm=V_{S,d-2} R_\pm^{d-2}$.
In the formulas above one has $l_P=1/M_P$, $M_P^{2-d}=G$ and $8\pi G=\kappa$.

\section{Thin-Wall Limit \label{sec:thw}}
In the thin-wall regime, when both vacua are nearly degenerate, the decay happens from vacuum to vacuum and the tunneling action can be expressed in terms of the wall tension $\sigma$ and the potential difference between the minima, $\Delta V\equiv V_\pp-V_\mm$. The $d=4$ derivation in the $V_t$ formalism, presented in \cite{E,Eg} both with and without gravity, is easy to generalize to $d>2$. 

The starting point is $|V_t'|\ll |(V-V_t)'|$. 
Without gravity, this implies that the EoM for $V_t$, (\ref{EoMVt}), gives 
$2(V-V_t)V_t''\simeq (V_t-V)'V_t'$, which holds for any $d$ and is integrated to give 
\be
V_t' \simeq -C \sqrt{2(V-V_t)} \ .
\label{Vtpthw}
\ee
From this approximate equality we can derive all thin-wall key relations. 
First, the integration constant $C$ can be related to the wall tension and the potential difference by integrating (\ref{Vtpthw}), noting that $\phi_0\simeq \phi_\mm$,  
\be
\sigma\equiv \int_{\phi_\pp}^{\phi_0}\sqrt{2(V-V_t)}\, d\phi
\simeq \frac{\Delta V}{C}\ .
\label{sigmaVt}
\ee

Second, if we plug (\ref{Vtpthw}) in the expression for $\rho$ [Eq.~(\ref{rho}), setting now $\kappa=0$], we get a constant value
\be
R_B=\frac{(d-1)}{C}=(d-1)\frac{\sigma}{\Delta V} \ .
\ee
This is the radius where all the field evolution of the bounce takes place, which is precisely the radius of the critical bubble. 
As one would expect, $R_B$ diverges for $\Delta V\to 0$.

Finally, plugging (\ref{Vtpthw})  in $s_0^{(d)}$ of (\ref{s0d}), one obtains an integrable action density leading to the thin-wall action
\be
S_{\rm thw}=\frac{(d-1)^{d-1}\pi^{d/2}}{\Gamma(1+d/2)}\frac{\sigma^d}{\Delta V^{d-1}}\ ,
\ee
which coincides with the thin-wall result obtained in \cite{Amariti}.\footnote{Notice that this action has the right dimensions, $[S_{\rm thw}]=0$, as $[V]=d$ and $[\sigma]=d-1$.} For $\Delta V\to 0$ the action diverges and tunneling becomes impossible.

With gravity, $|V_t'|\ll |(V-V_t)'|$ implies, from (\ref{EoMVtD}), $D_d'/D_d\simeq (V-V_t)'/[2(V-V_t)]$ which gives 
\be
D_d^2= V_t'{}^2+4\kappa_d(V-V_t)V_t\simeq 2C^2 (V-V_t)\ ,
\ee
from which 
\be
\sqrt{2(V-V_t)}\simeq \frac{|V_t'|}{\sqrt{C^2-2\kappa_dV_t}}\ .
\label{Vtpthwk}
\ee
This is the generalization of (\ref{Vtpthw}) with gravitational effects included. 
Now one has to pay attention to the type of false vacuum decay as this determines the sign of $V_t'$. Below we discuss the two qualitatively different cases separately. 

\subsection{Minkowski or AdS Vacua} 

For Minkowski or AdS false vacua, a thin-wall bounce typically exists when the false and true vacua are nearly degenerate or for sufficiently strong gravitational effects, such that the situation is close to critical (see discussion in Section~\ref{GQ}). For this type of vacua, $V_t(\phi)$ is monotonic with $V_t'\leq 0$.

Integrating (\ref{Vtpthwk}) in the interval $(\phi_\pp,\phi_0)\simeq(\phi_\pp,\phi_\mm)$, the wall tension is obtained, in terms of $C$, as
\be
\sigma = \left.\frac{1}{\kappa_d}\sqrt{C^2-2\kappa_d V(\phi)}\right|_{\phi_\pp}^{\phi_\mm}\ ,
\label{sigma}\ee
which can be solved for $C^2$ giving
\be
C^2=2\kappa_d V_\pp+\frac{1}{\sigma^2}\left(\Delta V-\delta V\right)^2\ ,
\label{C}
\ee
with
\be
\delta V\equiv\frac12 \kappa_d\sigma^2\ .
\ee
As in the case without gravity, if we plug the thin-wall relation $D_d=C\sqrt{2(V-V_t)}$ in the expression for $\rho$,  (\ref{rho}), we get
the radius of the thin-wall bounce (or of the critical nucleation bubble) as  
\be
R_B=\frac{(d-1)}{C}=
\left.\frac{(d-2)}{\kappa\sigma}\sqrt{1+\frac{R_B^2}{R(\phi)^2}}\right|_{\phi_\pp}^{\phi_\mm}\ .
\label{RB}
\ee
The limit $\Delta V\to 0$ is meaningless in this expression as it corresponds to a supercritical potential, with decay quenched by gravity. Below we discuss how the limit $R_B\to \infty$ is reached. 

Using the previous results we can express the action density $s^{(d)}$ in (\ref{sd}), with $V_t'\leq 0$, as a function of $V_t$ and $V_t'$ only that can be integrated exactly. The resulting thin-wall tunneling action is\footnote{Again the dimensions are correct, $[S_{\rm thw}]=0$, as $[C]=2$, $[\kappa]=2-d$ and $[z]=0$.}
\be
S_{\rm thw}^{\rm AdS}=\left.\frac{\pi^{d/2}R_B^{d-2}}{\kappa\Gamma(1+d/2)}
\left[d\sqrt{1-z}+(d-1)z\, {}_2F_1(1/2,d/2;d/2+1;z)\right]
\right|^{z_\pp}_{z_\mm}\ ,
\label{SthwR}
\ee
with
\be
z_\pm\equiv 2\kappa_d\frac{V_\pm}{C^2}=-\frac{R_B^2}{R_\pm^2}\ .
\ee
In appendix \ref{app:thwE} we show that this agrees with the Euclidean thin-wall action.

\begin{figure}[t!]
\begin{center}
\includegraphics[width=0.6\textwidth]{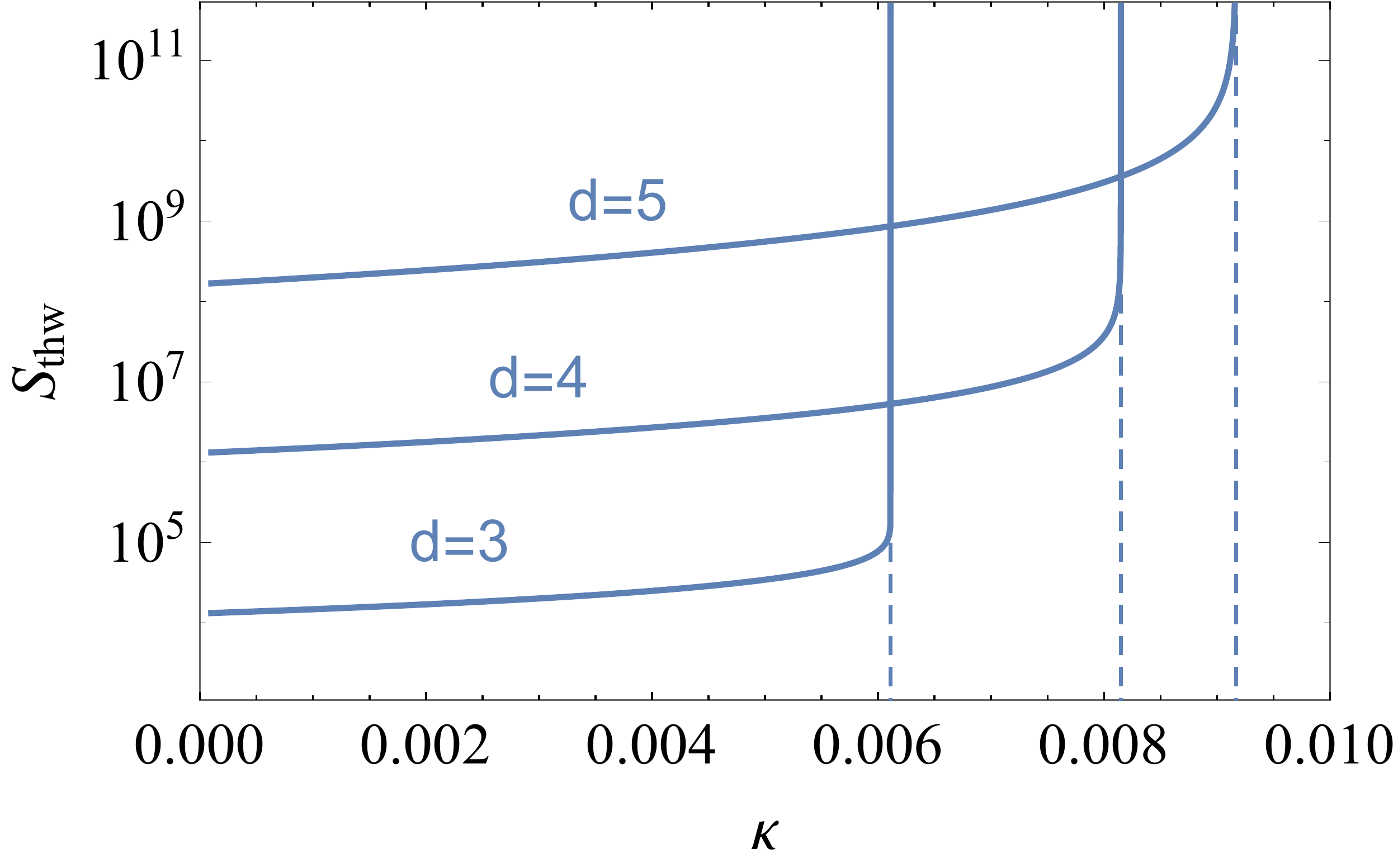}
\end{center}
\caption{Thin-wall tunneling action from (\ref{SthwR}) as a function of the strength of gravity, measured by a varying $\kappa$, for different spacetime dimensions, $d=3,4,5$, for $V_\pp=-0.1$, $V_\mm\simeq V_0=-0.5$ and $\sigma=5$. The dashed vertical lines show the critical $\kappa_c(d)$ values.
\label{fig:Sthw}
}
\end{figure}

Figure \ref{fig:Sthw} shows how this action depends on the strength of gravitational effects, measured by varying $\kappa$ and keeping other potential parameters fixed\footnote{Varying $\kappa$ is just a convenient way of exploring the effect of keeping $m_p$ fixed and varying any other mass scale $M$ in the problem (in effect varying $\kappa M^2$).}  as indicated, for different values of the spacetime dimension $d$. This figure illustrates several features, some of which have been discussed in previous subsections. First, for $\kappa=0$ we see that the action (and therefore the stability of the vacuum) increases with $d$. Each curve for fixed $d$ also shows how stronger gravity also tends to  stabilize the vacua. Eventually a critical value $\kappa_c(d)$ is reached where the action diverges. For $\kappa>\kappa_c(d)$ the vacuum is stable (gravitational quenching of the decay). The value of $\kappa_c(d)$ is simply obtained from the condition $D_d=0$, which in the thin-wall regime corresponds to $C^2=0$ (or $R_B=\infty$). This gives
\be
\kappa_c(d)=\frac{2(d-2)}{\sigma^2(d-1)}\left(\sqrt{-V_\pp}-\sqrt{-V_\mm}\right)^2\ .
\ee
Figure \ref{fig:Sthw} shows the values of $\kappa_c(d)$ as dashed lines. Finally, the figure also shows how changing $d$ at $\kappa=\kappa_c(d)$ transforms a critical case into subcritical, if $d$ is raised, or into supercritical, if $d$ is lowered.

\subsection{dS Vacua\label{sec:dSthw}}
For dS false vacua, instead, a thin-wall bounce can occur when the false and true vacua are nearly degenerate {\it and} gravitational effects are weak. (If they are not, then tunneling rates between nearly degenerate vacua are not supressed.)  In the thin-wall regime we now have $\phi_{0\pp}\simeq \phi_\pp$ and $\phi_{0}\simeq \phi_\mm$, so that tunneling occurs directly between the vacua. The $\phi_\mm$ vacuum can be lower (dS, Minkowski or AdS) or higher (dS) than the $\phi_\pp$ vacuum. 
We start discussing dS to dS transitions and will comment on decays from dS to Minkowski or AdS at the end. 

\begin{figure}[t!]
\begin{center}
\includegraphics[width=0.45\textwidth]{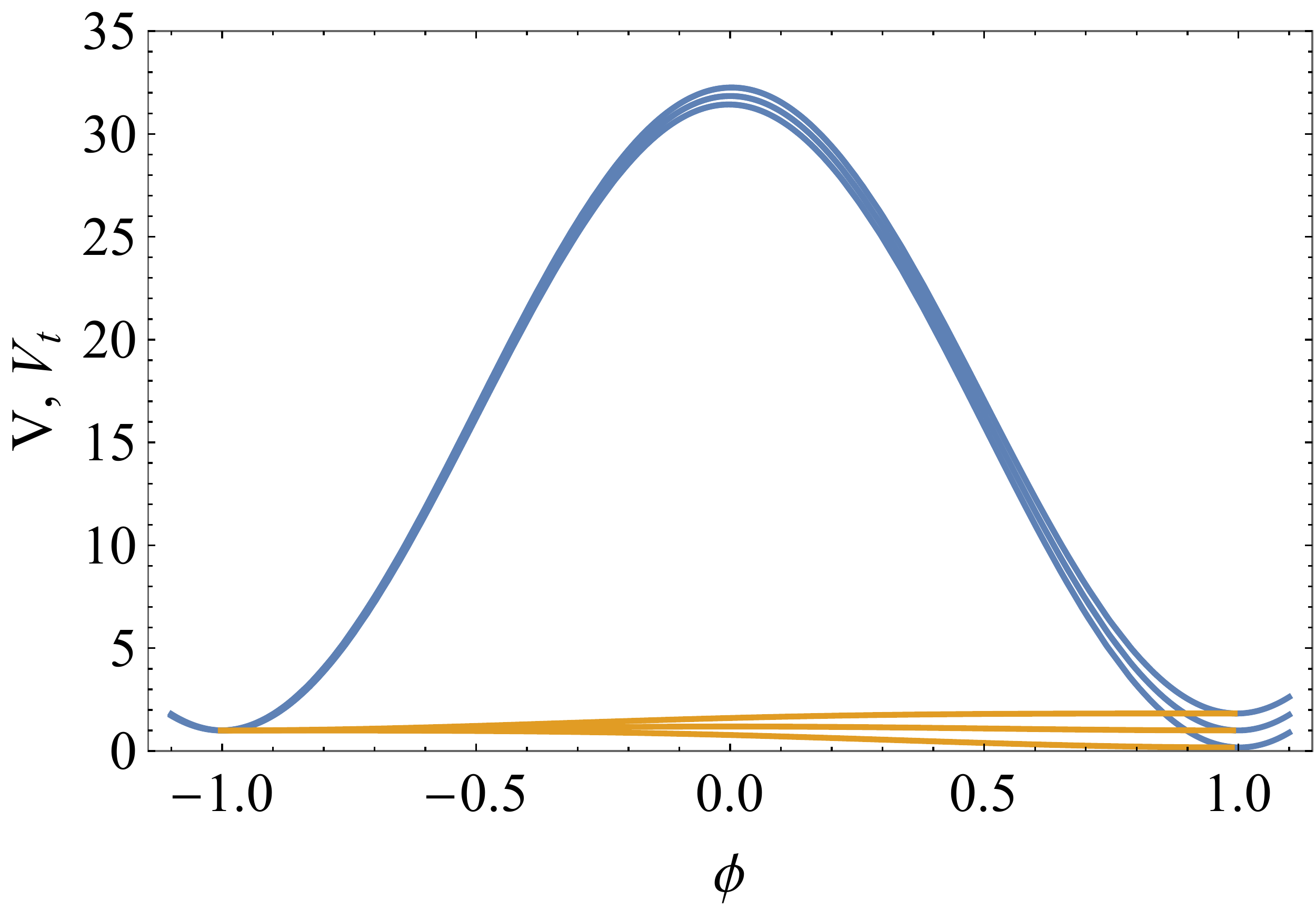}
\includegraphics[width=0.45\textwidth]{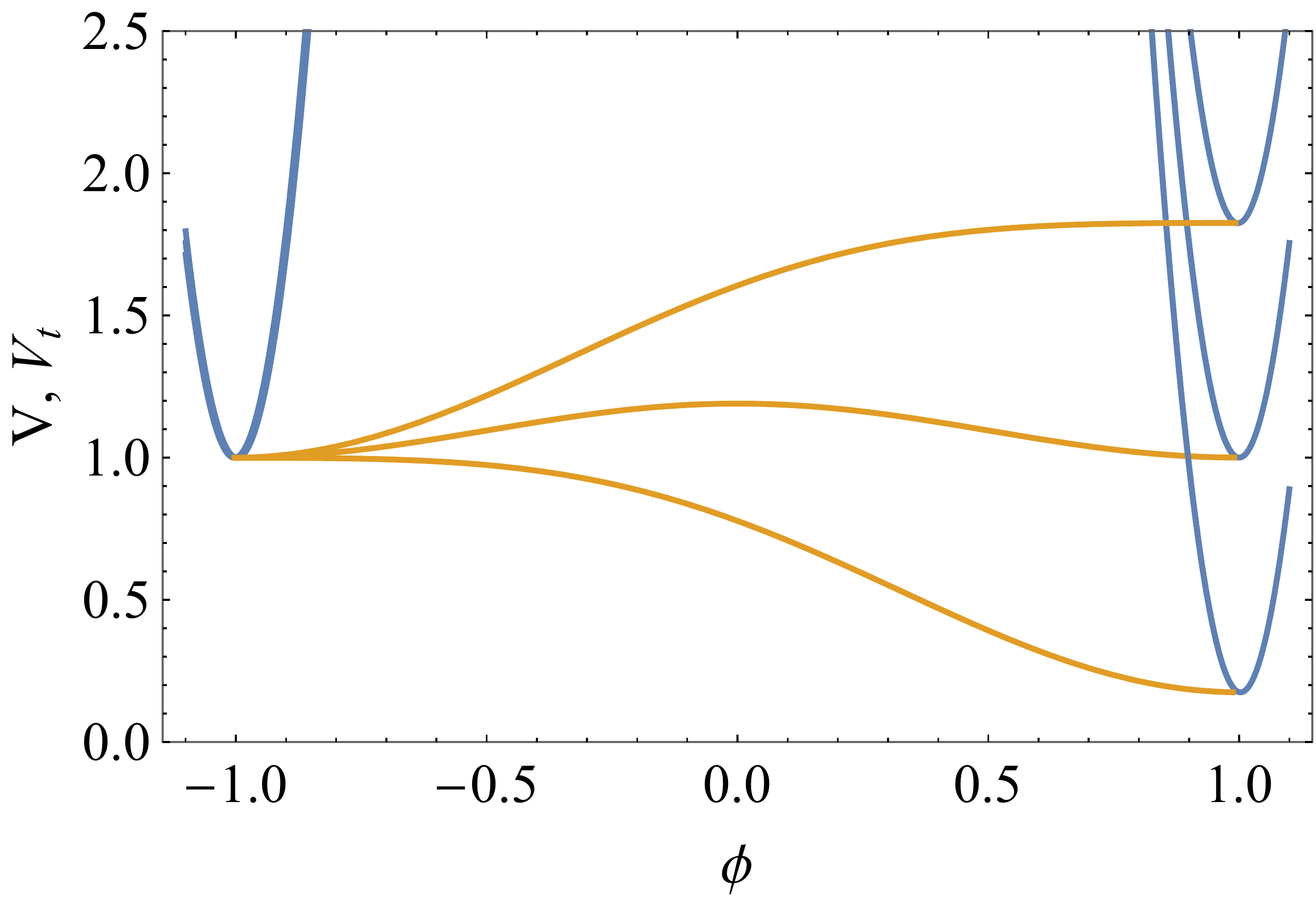}
\end{center}
\caption{Potentials and tunneling potentials for the three cases of dS decay: case (1) with $V_\mm<V_\pp-\delta V$ (lower curves); case (2) with $V_\mm>V_\pp+\delta V$ (upper curves) and  case (3) with $V_\mm=V_\pp$ (central curves). The right plot zooms on the left one to show the $V_t$ structure.
\label{fig:dSCasesVt}
}
\end{figure}

For dS decays, $V_t(\phi)$ first grows and then decreases. Let us call $\phi_T$ the field value at which the maximum occurs, with $V_T\equiv V_t(\phi_T)$, $V_t'(\phi_T)=0$. 
From (\ref{Vtpthwk}) we see that there are three different possibilities for the location of $\phi_T$: (1)
$\phi_T\simeq \phi_\pp$; (2) $\phi_T\simeq \phi_\mm$ or (3) $\phi_T$ is somewhere at the wall. In cases (1,2) $V_t(\phi)$ is well approximated by a monotonic function between the minima, and therefore case (1) requires $V_\pp>V_\mm$ while case (2) needs $V_\pp<V_\mm$. In case (3), $V_\pp\simeq V_\mm$ (we make this more precise below) and  (\ref{Vtpthwk}) gives $\sqrt{C^2-2\kappa_d V_T}=0$. Figure \ref{fig:dSCasesVt} illustrates the shape of $V_t$ for these three cases using some numerical solutions for $V_t$.

The integral of (\ref{Vtpthwk}) across the wall now gives
\be
\sigma = \frac{1}{\kappa_d}\left[\sqrt{C^2-2\kappa_d V_\pp}+\sqrt{C^2-2\kappa_d V_\mm}-2\sqrt{C^2-2\kappa_d V_T}
\right]\ ,
\label{sigmadS}
\ee
which is valid for the three different cases using $V_T=V_\pp$ for (1); $V_T=V_\mm$ for (2); and  $\sqrt{C^2-2\kappa_d V_T}=0$ for (3). If we solve for $C^2$ we get (\ref{C}) again for all three cases but we also learn that the solution is valid for case (1) if $V_\pp>V_\mm-\delta V$; for case (2) if $V_\mm>V_\pp+\delta V$
and for case (3) if $|V_\pp-V_\mm|<\delta V$.

Plugging $D_d\simeq C\sqrt{2(V-V_t)}$ in (\ref{rho}), the radius
of the critical bubble is again related to $C$ by $R_B=(d-1)/C$. Substituting $C$ in terms of $R_B$ in (\ref{sigmadS}) one can solve for $R_B$, which turns out to be given by the same formula in all three cases
\be
\frac{1}{R_B^2}=\frac{1}{R_\pp^2}+\frac{\left(\Delta V-\delta V\right)^2}{(d-1)^2\sigma^2}=\frac{1}{R_\mm^2}+\frac{\left(\Delta V+\delta V\right)^2}{(d-1)^2\sigma^2}
\label{RBdS}
\ .
\ee
Note, in particular, that for $\Delta V=0$ one has $R_B=R_\pp/\sqrt{1+R_\pp^2\kappa^2\sigma^2/[4(d-2)^2]}$, a finite value (in contrast with the case of Minkowski or AdS vacua, for which $\Delta V\to 0$ corresponds to a stable vacuum).

Using the thin-wall relations presented above, the tunneling action density can be integrated exactly. Paying attention to the sign of $V_t'$ one gets the action
\bea
S_{\rm thw}&=&
\frac{\pi^{d/2}R_B^{d-2}}{\kappa \Gamma(1+d/2)}\left\{\left.\left[d\sqrt{1-z}+(d-1)z\, {}_2F_1(1/2,d/2,d/2+1,z)\right]
\right|_{z_\mm}^{z_T}\right.\nonumber\\
&+&\left.\left[d\sqrt{1-z}+(d-1)z {}_2F_1(1/2,d/2,d/2+1,z)\right]\right|_{z_\pp}^{z_T} \nonumber\\
&+&\left. 2(d-1)\, {}_2F_1(1/2,d/2,d/2+1,1)\left.\left[z^{1-d/2}\right]\right|_{z_T}^{z_\pp}\right\}\ ,
\label{SVTthw}
\eea
where now
\be
z_\pm\equiv 2\kappa_d\frac{V_\pm}{C^2}=\frac{R_B^2}{R_\pm^2}\ , 
\ee
while $z_T,R_T$  take different values for each case: for (1), $T=+$; for (2) $T=-$, and for (3)
\be
z_T=1\ ,  \quad V_T = \frac{C^2}{2\kappa_d}\ ,\quad R_T^2\equiv \frac{(d-1)(d-2)}{2\kappa V_T}\ .
\ee 
In this last case, the $T$-dependent terms in (\ref{SVTthw}) cancel out. The height of the bump in $V_t$ is
\be
V_T-V_\pp = \frac{(\Delta V-\delta V)^2}{4\delta V}\ .
\ee

\begin{figure}[t!]
\begin{center}
\includegraphics[width=0.6\textwidth]{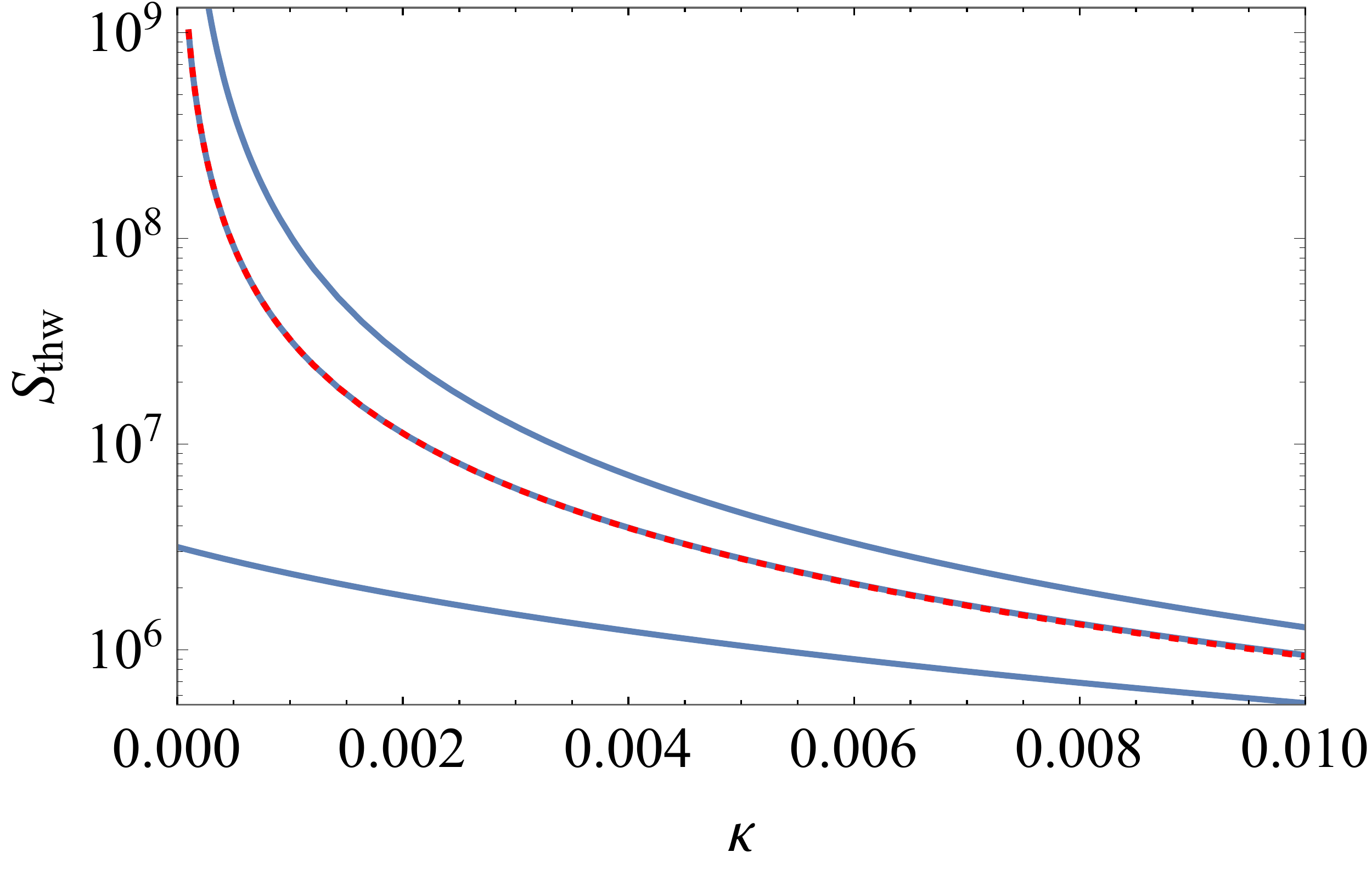}
\end{center}
\caption{Thin-wall tunneling action from (\ref{SVTthw}) as a function of the strength of gravity, measured by a varying $\kappa$, for the three different types of dS decay: case (1) with $V_\mm<V_\pp-\delta V$ (lower curve); case (2) with $V_\mm>V_\pp+\delta V$ (upper curve) and  case (3) with $V_\mm=V_\pp$ (central curve). 
The red dotted line shows the approximation (\ref{Sthwapp}) for the $V_\mm=V_\pp$ case.
\label{fig:SthwdS}}
\end{figure}

Figure~\ref{fig:SthwdS} shows the thin-wall action for the three types of dS decay as a function of $\kappa$, taking $d=4$, $V_\pp=1$, $\sigma=10$ and $V_\pp=\{0.25,1.75,1\}$ for types (1), (2) and (3) respectively.
The thin-wall action grows now with decreasing $\kappa$, as expected, and, for the special case $\Delta V=0$ the small $\kappa$ expansion of the action (\ref{SVTthw}), which reads 
\be
S_{thw}\simeq V_{S,d-1} R_\pp^{d-1}\sigma\left[1-\frac{(d-1)\kappa_d\sigma^2}{48 V_\pp}\right]\ ,
\quad ({\rm for}\;\Delta V=0)\ ,
\label{Sthwapp}
\ee
gives an excellent approximation (see red dotted line in the figure).
For cases (2) and (3) the action diverges for $\kappa\to 0$, as gravity is needed to tunnel up or to the same level. Instead, for case (1) the action is finite for $\kappa\to 0$ as down-tunneling is still possible without gravity.

Finally, if the false dS vacuum decays into Minkowski, the thin-wall action is simply the $R_\mm\to\infty$ limit of (\ref{SVTthw}), which gives, for case (1) decays, with $V_\pp>\delta V$,
\be
S_{{\rm thw},(1)}^{{\rm dS}\to{\rm Mink}}=
\frac{\pi^{d/2}R_B^{d-2}}{\kappa \Gamma(1+d/2)}\left[d\left(\sqrt{1-z_\pp}-1\right)+(d-1)z_\pp\, {}_2F_1(1/2,d/2,d/2+1,z_\pp)\right]\ ,
\ee
and, for case (3) decays, with $V_\pp>\delta V$,
\bea
S_{{\rm thw},(3)}^{{\rm dS}\to{\rm Mink}}&=&
\frac{\pi^{d/2}R_B^{d-2}}{\kappa \Gamma(1+d/2)}\left[
2(d-1) z_\pp^{1-d/2}{}_2F_1(1/2,d/2,d/2+1,1) 
\right.\nonumber\\
&-&\left.
(d-1)z_\pp\, {}_2F_1(1/2,d/2,d/2+1,z_\pp)
-d\left(\sqrt{1-z_\pp}+1\right)
\right]\ .
\eea

If the dS vacuum decays into AdS, the thin-wall action is obtained
from (\ref{SVTthw}) and (\ref{C}) simply replacing $R_\mm^2\to-R_\mm^2<0$.
As in the case for dS to Minkowski, the decay can be of type (1) or type (3) depending on ${\rm sign}(\Delta V-\delta V)$.
\section{Exactly Solvable Models \label{sec:exact}} 
\noindent

It has been illustrated in several papers \cite{E,Eg,EFH} how the $V_t$ formulation can be used to obtain potentials with exactly solvable false vacuum decay.  The procedure is to postulate a simple $V_t$ and solve its EoM for $V$, which is simpler that solving for $V_t$ given $V$, as the EoM is a first-order differential equation for $V$. With gravity, this procedure was detailed in \cite{EFH}, which was restricted to $d=4$. We generalize the method to arbitrary $d>2$ in this section.\footnote{Analytical potentials (or bounces) for general $d$ have been obtained before, see {\it e.g.} \cite{Amariti,Miha} for examples (without gravity).}

Starting with the expression for $D_d$ in (\ref{Dd}) we can write $V(\phi)$ in terms of $V_t(\phi)$ and $D_d(\phi)$ as
\be
V(\phi)=V_t+\frac{D_d^2-V_t'{}^2}{4\kappa_d V_t}\ .
\label{VfromD}
\ee
Using (\ref{VfromD}), the EoM (\ref{EoMVt}), given in terms of $V_t$ and $V$, can be rewritten in terms of $D_d$ and $V_t$ as
\be
 V_t''+\frac{2\kappa}{d-2} V_t+\frac{D_d^2-V_t'{}^2}{2V_t}- V_t'\frac{D_d'}{D_d}=0\ .
\label{EoMD}
\ee
This  equation can be integrated to obtain
$D_d^2$ in terms of $V_t$ formally as
\be
D_d^2(\phi)= \frac{V_t'{}^2}{1-V_{t}F}\ ,
\label{D2F}
\ee 
where
\be
F(\phi)\equiv \frac{4\kappa}{(d-2)E(\phi)} \int_{\phi_0}^{\phi}\frac{E(\tilde\phi)}{V_t'(\tilde\phi)}d\tilde\phi\ ,\quad
E(\phi)\equiv \exp\left[\frac{4\kappa}{(d-2)}\int_{\phi_0}^{\phi}\frac{V_t(\tilde\phi)}{V_t'(\tilde\phi)}d\tilde\phi\right]\ ,
\label{FE}
\ee
and $\phi_0$ is a reference field value, that we take to be one of the two contact points between $V$ and $V_t$, so that $D_d^2(\phi_0)=V_t'{}^2(\phi_0)$.

If $V_t$ is simple enough for the integrals $E$ and $F$ to be performed analytically we obtain an explicit $V$ for which vacuum decay is under analytic control. Having found $D_d^2$, we  plug it in (\ref{VfromD}) and obtain $V$ as\footnote{Without gravity, we get  $V(\phi)=V_t(\phi)+\frac{V_t'(\phi)^2}{d-1}\int_{\phi_0}^\phi d\bar\phi/V_t'(\bar\phi)$.}
\be
V(\phi)=V_t+\frac{V_t'{}^2}{4\kappa_d(1/F-V_t)}\ .
\label{formalV}
\ee
We can recast (\ref{EoMD}) as a
differential equation for $F$ as
\be
F'V_t'=\frac{4\kappa}{d-2}(1-FV_t)\ ,
\label{EoMF}
\ee
and it is also possible to get analytic examples of $V$ by postulating appropriately a simple $F$. 
As in the $d=4$ case, the results above are general and apply to decays from Minkowski, AdS or dS vacua. The procedure can be used to obtain analytical potentials in all these cases. 

The results just obtained for general $d$ are slight modifications
of the $d=4$ case, with a simple rescaling of $\kappa$. The general strategy discussed in \cite{EFH} can thus be applied immediately to the general $d$ case and the interested reader is directed to that paper for further details.  In fact, it is straightforward to generalize many analytic examples found in \cite{EFH} to general $d$. Here we simply do this for the simplest example of \cite{EFH}, which follows from assuming $F=a V_t +b$ and fixing $a$, $b$ and an integration constant appropriately. One gets
\be
V_t(\phi)=\frac{2}{\sin^2\theta}\left[\cos\theta+\cos\left(2\sqrt{\frac{\kappa}{d-2}}\phi\right)\right]\ ,
\ee
where $\theta$ is a free parameter, with $0<\theta<\pi/2$,
and 
\be
V(\phi)=\frac{2}{(d-1)\sin^2\theta}\left[(d-2)\cos\theta+d\cos\left(2\sqrt{\frac{\kappa}{d-2}}\phi\right)\right]\ .
\ee
This solution describes the CdL instanton part of a dS to dS transition, which exists in the interval $\phi\in(-\alpha,\alpha)$ where $\alpha\equiv \theta\sqrt{(d-2)/(4\kappa)}$.

As done in \cite{EFH} for $d=4$, one can derive the field and metric Euclidean profiles, $\phi(\xi)$ and $\rho(\xi)$, now for general $d$. By integrating $d\phi/d\xi=-\sqrt{2(V-V_t)}$, one gets
\be
\phi(\xi)=-\sqrt{\frac{d-2}{\kappa}}\ \mathrm{am}\left(\left.\sqrt{\frac{2\kappa}{(d-2)(d-1)}}\,\frac{(\xi-\xi_e/2)}{\cos(\theta/2)}\right|\csc^2(\theta/2)\right)
\ee
where $\mathrm{ am}(u|m)$ is the Jacobi amplitude function
and 
\be
\xi_e\equiv\sqrt{\frac{(d-2)(d-1)}{2\kappa}}\sin\theta\ K(\sin^2(\theta/2))\ ,
\ee
with $K(m)$ the complete elliptic function of the first kind.  The metric function can be obtained from $\rho=(d-1)\sqrt{2(V-V_t)}/D_d$, see (\ref{rho}), as
\be
\rho(\xi)=\frac12\sqrt{\frac{(d-2)(d-1)}{\kappa}\left[\cos\left(2\sqrt{\frac{\kappa}{d-2}}\phi(\xi)\right)-\cos\theta\right]}\ .
\ee
The CdL instanton is defined in the interval $\xi\in (0,\xi_e)$, with $\rho(0)=\rho(\xi_e)=0$.

\section*{Acknowledgments} 
The work of J.R.E. has been supported by the following grants: IFT Centro de Excelencia Severo Ochoa SEV-2016-0597, CEX2020-001007-S and by PID2019-110058GB-C22 funded by MCIN/AEI/10.13039/501100011033 and by ``ERDF A way of making Europe''. The work of J.F.F. is supported by NSERC.

\appendix

\section{Action from Canonical Transformation\label{app:cantransf}}

To obtain the action (\ref{sd}) using a canonical transformation from the Euclidean result we follow closely \cite{EJK}, generalizing to $d>2$.

\subsection{AdS or Minkowski Vacua}

First one rewrites the Euclidean action (\ref{SE}) in terms of 
$\xi(\phi)$ and $\rho(\phi)$, as
\be
S_E =\int_{\phi_+}^{\phi_0}L\, d\phi= -V_{S,d-1} \int_{\phi_+}^{\phi_0}
\left[ \rho^{d-1} \left( \frac1{2 \xi'^2} + V\right)
- \frac{(d-1)(d-2)\rho^{d-3}}{2\kappa} \left( \frac{\rho'^2}{\xi'^2} + 1 \right) \right] \, \xi' \, d\phi \, .
\ee
The canonical momenta are 
\be
\label{prho1}
p_\rho = \frac{2(d-1)(d-2) \pi^{d/2} \rho^{d-3} \rho'}{\kappa\Gamma(d/2) \xi'} \, ,
\ee
and
\be
p_\xi = \frac{\pi^{d/2} \rho^{d-3}}{\kappa \Gamma(d/2)\xi'^2} 
\left[ (d-1)(d-2)(\xi'^2-\rho'^2) + \kappa \rho^2 (1-2 V \xi'^2) \right] \, ,
\ee
and the Hamiltonian can be written as
\be
H = \frac{2 \pi^{d/2} \rho^{d-1}}{\Gamma(d/2)\xi'} \, ,
\label{Hgrav}
\ee
where $\xi'$ is understood to be a function of $\xi$, $\rho$ and 
the canonical momenta $p_\xi$ and $p_\rho$. In fact, the Hamiltonian does not depend on $\xi$, and $p_\xi$ is constant which is 0 from (\ref{CdL2}), see \cite{EJK}. One gets
\be
H = -\frac{2\pi^{d/2}\rho^{d-1}}{\Gamma(d/2)} \sqrt{2V-\frac{(d-1)(d-2)}{\kappa\rho^2}+\frac{\kappa \Gamma(d/2)^2p_\rho^2\rho^{4-2d}}{4\pi^d(d-1)(d-2)}} \, .
\ee

Now we replace $\rho$ and its momentum $p_\rho$ by $V_t$ and its momentum $P$, with
\be
V_t = V- \frac{1}{2 \xi'^2} = \frac{(d-1)(d-2)}{2\kappa\rho^2}+\frac{\kappa \Gamma(d/2)^2p_\rho^2\rho^{4-2d}}{8\pi^d(d-1)(d-2)}\, ,
\label{Vtgrav}
\ee
so that $p_\rho$ is
\be
\label{prho2}
p_\rho = \frac{2 \pi^{d/2}(d-1)(d-2) \rho^{d-3}}{\kappa\Gamma(d/2)} \sqrt{1 -\frac{2 V_t \kappa \rho^2}{(d-1)(d-2)}} \, .
\ee
The generating function is obtained by integrating $p_\rho=\partial G/\partial\rho$ and takes the form
\be
G = \frac{2(d-1)\pi^{d/2} \rho^{d-2}}{\kappa\Gamma(d/2)}{}_2F_1\left(-\frac12,\frac{d}{2}-1,\frac{d}{2},\frac{2\kappa\rho^2V_t}{(d-1)(d-2)}\right) \, .
\label{G}
\ee
The canonical momentum is then obtained by $P = -\partial G/\partial V_t$.
The transformed Hamiltonian reads
\be
K = -\frac{2 \pi^{d/2}}{\Gamma(d/2)} \rho^{d-1} \sqrt{2(V - V_t)} \, ,
\label{K}
\ee
where $\rho$ can in principle be expressed as a function of $V_t$ and $P$. To get the transformed Lagrangian, $L=PV_t'-K$, we need
$V_t'$, which we can get directly from $K$ as
\be
V_t' = \frac{\partial K}{\partial P} = \frac{3 K}{\rho} \left(\frac{\partial P}{\partial \rho}\right)^{-1}
= -\frac{(d-1)}{\rho} \sqrt{2(V-V_t)} \sqrt{1-\frac{2 V_t \kappa \rho^2}{(d-1)(d-2)}} \, ,
\ee
from which we also obtain $D_d^2 \equiv V_t'^2 + 4\kappa_d (V - V_t) V_t= 2(d-1)^2(V-V_t)/\rho^2$.

The transformed Lagrangian (action density) then takes the form
\bea
L &=& \frac{(d-2)\pi^{d/2}D_d}{\kappa\Gamma(d/2)V_t}\left[\frac{(d-1)\sqrt{2(V-V_t)}}{D_d}\right]^{d-2}\nonumber\\
&&\times\left[1+\frac{(d-2)V_t'^2}{D_d^2}+(d-1)\frac{V_t'}{D_d}{}_2F_1\left(-\frac12,\frac{d}{2}-1,\frac{d}{2},1-\frac{(V_t')^2}{D_d^2}\right)
\right]\, .
\label{LAdS}
\eea
To get the final tunneling action we need to subtract the false vacuum background action and a boundary term from $G$
\be
S= \int_\pp^0L\, d\phi+\left. G \right |_\pp^0 - S_{E+}\ ,
\ee
where we use the subindex $0$  for $\phi=\phi_0$ or $\xi=0$ and the subindex $+$ for $\phi=\phi_\pp$ or $\xi=\infty$.
The boundary piece diverges and is
\be
\left. G \right |_+^0 = -G(\rho_\infty,V_\pp)\ ,
\ee
where $\rho_\infty\to\infty$.
The false vacuum background piece is also divergent and given by
\be
S_{E+}=-\frac{2(d-1)(d-2)\pi^{d/2}}{\kappa\Gamma(d/2)}\int_0^{\rho_\infty}\rho^{d-3}\sqrt{1-\frac{2\kappa V_\pp \rho^2}{(d-1)(d-2)}}d\rho=- G(\rho_\infty,V_\pp)\ ,
\ee
so that both terms cancel and (\ref{LAdS}) gives in fact the final result.
Indeed this is in agreement with the action in (\ref{sd}) with $V_t'\leq 0$ or (\ref{SminkAdS}) as can be shown using  hypergeometric function identities.

\subsection{dS Vacua}

The previous derivation goes through for dS vacua as well, paying attention to two facts: the canonical transformation holds for the ``CdL regime'' of the bounce (when $V_t\neq V$) and one has to keep track of the sign change of $\dot\rho$ (at $\xi_T$) or $V_t'$ (at $\phi_T$), as usual. Equation (\ref{prho1}) tells us that $p_\rho$ also flips sign and, for dS (\ref{prho2}) should be multiplied by ${\rm sign} {\dot\rho}=-{\rm sign} {V_t'}$. Still, $p_\rho$ is continuous since the square root vanishes at $\xi_T$. The generating function $G_{dS}$ is sensitive to the sign flip from the integrand $p_\rho$ so that $G_{dS}(\xi\leq \xi_T)=G_{dS}(\phi_T\leq \phi\leq\phi_0)=G$ and $G_{dS}(\xi<\xi_T)=G_{dS}(\phi_{0\pp}\leq \phi\leq\phi_T)=-G+2G(\phi_T)$ with $G$ as given in (\ref{G}). One also has $dG_{dS}/d\phi=0$ at $\phi_T$. On the other hand, $K$ does not change as it depends on $(V_t')^2$ while  $L$ does change, with the replacement $V_t'\to -|V_t'|$ in the term linear in $V_t'$.  

The boundary term from $G_{dS}$ gives a contribution to the tunneling action
\bea
\int_{\phi_{0\pp}}^{\phi_0} \frac{dG_{dS}}{d\phi} d\phi  
= G(\phi_{0\pp})+G(\phi_0)-2G(\phi_T)=\frac{-4\pi^{(d+1)/2}}{\kappa\Gamma((d-1)/2)}R_T^{d-2}\ .
\eea
Here $G(\phi_{0\pp})=G(\phi_0)=0$ as $\rho=0$ at both points
and $G(\phi_T)$ is simplified by the fact that the argument of the hypergeometric function is 1 at $\phi_T$.

The false vacuum background contribution to be subtracted is as calculated in (\ref{SEp}). Combining the three pieces just discussed one gets precisely the action (\ref{SdSalt}), after using  hypergeometric identities.

\section{Action in Terms of Elementary Functions\label{app:elementary}}

The action for the tunneling potential in general $d$ dimensions derived in the text can be expressed in terms of elementary functions, as has been illustrated by the $d=3,4,5$ cases. Here we
give the expression for general $d$, which can be obtained by exploiting recursively the relation
\be
F^{(d)}(z)=\frac{d}{(d-3)z}\left[
(1-z)^{(3-d)/2}-F^{(d-2)}(z)
\right]\ ,
\ee
for
\be
F^{(d)}(z)\equiv {}_2F_1((d-1)/2,d/2;1+d/2;z)\ .
\ee
For even $d$, the relation above can be used to relate $F^{(d)}(z)$
to
\be
F^{(2)}(z)=\frac{2}{z}\left(1-\sqrt{1-z}\right)\ .
\ee
For odd $d$, the relation above can be used to relate $F^{(d)}(z)$
to
\be
F^{(3)}(z)=\frac{3}{z^{3/2}}\left({\rm arctanh}\sqrt{z}-\sqrt{z}\right)\ .
\ee
In this way we get
\be
F^{(2n)}(z)=\frac{-(2n)!!}{(2n-3)!!}\left[
\sum_{k=1}^{n-1}\frac{(2n-2k-3)!!(-1)^k}{(2n-2k)!!z^k(1-z)^{n-k-1/2}}+\frac{(-1)^n}{z^{n}}-\frac{(-1)^n}{z^{n}(1-z)^{-1/2}}\right]\ ,
\ee
and
\be
F^{(2n+1)}(z)=\frac{-(2n+1)!!}{(2n-2)!!}\left[
\sum_{k=1}^{n-1}\frac{(2n-2k-2)!!(-1)^k}{(2n+1-2k)!!z^k(1-z)^{n-k}}-\frac{(-1)^n}{z^{n}}+\frac{(-1)^n}{z^{n+1/2}}{\rm arctanh}\sqrt{z}\right]\ .
\ee
These results can then be plugged in the expression for the tunneling action $s^{(d)}$ in (\ref{sd}).

\section{Equivalence with Euclidean Action\label{app:equiv}}
To show the equivalence of the $S[V_t]$ action obtained in section~\ref{sec:action} with the standard Euclidean action $S_E[\phi,\rho]$, we rewrite the latter as a field-space integral of the action density in terms of $V_t$ quantities. Before doing that, it is convenient to first rewrite the Euclidean action in a simpler manner, which can be done  as follows.
Taking a further $\xi$ derivative of (\ref{CdL2}) we have
\be
\ddot\rho=-\frac{\kappa \rho}{(d-1)}\left(\dot\phi^2+\frac{2}{d-2}V\right) \ .
\ee
Using this in (\ref{R}) we get 
\be
R=2\kappa \left(\frac12 \dot\phi^2+\frac{d}{d-2}V\right) .
\ee
Plugging this in (\ref{SE0}) we arrive at
\be
S_E[\phi,\rho]=-\frac{2V_{S,d-1}}{d-2}\int_0^{\xi_e}
d\xi \rho^{d-1}V(\phi)+S_{\rm GHY}\ .
\label{SEs}
\ee
The Euclidean tunneling action is the difference between the action of the CdL instanton and the background false vacuum action
\be
S_E[\phi_\pp,\rho_\pp]\equiv S_{E,+}=-\frac{2V_{S,d-1}}{d-2}\int_0^{\xi_{e,+}}
d\xi_\pp \rho_\pp^{d-1}V_\pp+S_{\rm GHY,+}\ .
\ee
Here we call the radial coordinate $\xi_\pp$ to distinguish it from the CdL $\xi$ coordinate, see below. 

\subsection{AdS or Minkowski Vacua} 

Consider first the case of AdS or Minkowski false vacua, for which $\xi_e=\xi_{e,+}=\infty$, causing both $S_E[\phi,\rho]$ and $S_E[\phi_\pp,\rho_\pp]$ to diverge at $\xi,\xi_\pp\to\infty$, while their difference $\Delta S_E$ is finite.
A convenient way to control this cancellation of divergences is to rewrite $S_E[\phi_\pp,\rho_\pp]$ as a $\xi$ integral that can be combined with $S_E[\phi,\rho]$ to get an integral of a finite action density.
The false vacuum metric function is $\rho_\pp(\xi_\pp)=R_\pp \sinh(\xi_\pp/R_\pp)$.
In the Minkowski limit ($V_\pp\to 0$) one recovers simply $\rho_\pp=\xi_\pp$. To rewrite $S_{E,+}$ as a $\xi$ integral, we simply map $\xi_\pp$ to $\xi$ by imposing the relation $\rho_\pp(\xi_\pp)=\rho(\xi)$, which we can do as both functions grow monotonically from 0 to $\infty$ over the same $(0,\infty)$ interval. Then, the $\xi$ derivative of this relation gives
\be
d\xi_\pp =  \frac{\dot\rho}{\sqrt{1+\rho^2/R_\pp^2}}\, d\xi\ .
\ee
Notice that for $\xi\to \infty$ we have $\phi(\xi)\to\phi_\pp$ and $\dot\phi(\xi)\to 0$ implying, from (\ref{CdL2}), that $d\xi/d\xi_\pp\to 1$ and therefore
\be
S_{\rm GHY}-S_{{\rm GHY},+}=-\frac{(d-1)V_{S,d-1}}{\kappa}\left.\rho^{d-2}\dot\rho\left(1-\frac{d\xi}{d\xi_\pp}\right)\right|_{\xi=0}^{\xi=\infty}=0\ .
\ee
Putting all the pieces together, we get
\be
\Delta S_E = -\frac{4\pi^{d/2}}{(d-2)\Gamma(d/2)}\int_0^\infty \rho^{d-1}\left[V-\frac{\dot\rho V_\pp}{\sqrt{1+\rho^2/R_\pp^2}}\right]d\xi\ .
\ee
We can then rewrite this as a $\phi$ integral (with $\phi_\pp=0$)
\be
\Delta S_E =\int_0^{\phi_0}s_E^{(d)}d\phi\ ,
\ee
of a $V_t$-dependent action density by using the relations between Euclidean formalism and $V_t$ formalism derived in section \ref{sec:action}. We finally obtain the Euclidean density for the case of AdS or Minkowski vacuum decay as
\be
{\it s}_E^{(d)}= -\frac{4\pi^{d/2}(d-1)^{d-1}}{(d-2)\Gamma(d/2)}\frac{\left[2(V-V_t)\right]^{d/2-1}}{D_d^{d-1}}\left(V+\frac{\Vp V_t'}{\Ddp}\right)\ ,
\ee
where
\be
\Ddp \equiv \sqrt{V_t'{}^2+4\kappa_d (V-V_t)(V_t-\Vp)}\ .
\ee
The relation between the action densities in Euclidean and $V_t$ formulations is 
\be
{\it s}^{(d)}-{\it s}^{(d)}_E + G\  {\rm EoM} = \frac{dH}{d\phi}\ ,
\label{diffs}
\ee
where $ {\rm EoM}$ is the ``equation of motion'' for $V_t$, given by the LHS of (\ref{EoMVt}), and
\be
G=-\frac{2(2\pi)^{d/2}(d-1)^{d}}{(d-2)\Gamma(d/2)}\frac{(V-V_t)^{d/2-1}}{D_d^{d+1}}\left(V_t+\frac{\Vp V_t'}{\Ddp}\right)\ ,
\ee
while\footnote{To find $H(\phi)$  we used the homotopy operator method (see {\it e.g.} \cite{exactd}).}
\bea
H&=&
\frac{2(d-1)^d\left[2\pi(V-V_t)\right]^{d/2}}{(d-2)\Gamma(d/2+1) D_d^{d-1}}\left[\frac{\Vp}{\Ddp}\, _2F_1\left(1/2,1;d/2+1;1-D_d^2/\Ddp^2\right)\right.\nonumber\\
&+&\left.\frac{V_t}{V_t'}\, _2F_1\left(1/2,1;d/2+1;1-D_d^2/V_t'{}^2\right)\right]
-\frac{2\pi^{(d+1)/2}R_t^{d-2}}{\kappa \Gamma[(d-1)/2]}
\left(1+\frac{V_t'}{|V_t'|}\right)\ .
\label{H}
\eea
Equation (\ref{diffs}) holds for $dH/d\phi$ evaluated on-shell, {\it i.e.} with $V_t''$ as determined by the EoM for $V_t$.
 As a cross-check, the case $d=4$ reproduces the function
found in \cite{Eg}.

Integrating 
(\ref{diffs}) in $\phi$ one gets
\be
 S - \Delta S_E  = H(\phi_0) - H(\phi_\pp)\ .
\ee
To prove $S = \Delta S_E$ we should then check that $H(\phi)$ vanishes at the boundaries. Noting that $V_t'(\phi_0)<0$ and $D_d(\phi_0)=D_{d\pp}(\phi_0)=-V_t'(\phi_0)$, it follows that $H(\phi_0)=0$. To prove $H(\phi_\pp)=0$ is a bit more laborious. One has $D_d(\phi_\pp)=D_{d\pp}(\phi_\pp)=0$ 
[with $V'(\phi_\pp)=0$ at the false vacuum] and one needs the ratios of these quantities, so one needs to know in more detail how $V_t',D$ and $D_\pp$ approach zero. 

Consider the AdS case ($V_\pp<0$) first. Near the false vacuum, $\phi_\pp=0$, we can approximate the potential by keeping up to its second derivative
\be
V(\phi) = V_\pp + \frac12 m^2 \phi^2 +\dots
\ee
Solving the equation of motion (\ref{EoMVt}) for the above potential 
leads to the expansion for the tunneling potential
\be
V_t(\phi) = V_\pp + \frac12 B \phi^2 + B_\alpha \phi^{2+\alpha}+\dots
\ee
with 
\be
B=\kappa_d V_\pp\left(1+\sqrt{1-\frac{2m^2}{\kappa_d V_\pp}}\right)<0\, ,
\label{B}
\ee
and
\be
\alpha = \frac{4\kappa V_\pp}{(d-2)B}>0\ .
\label{a}
\ee
$B_\alpha$ is a free constant fixed by
the boundary condition at $\phi_0$. 
From this result it follows that 
\be
V_t',D_{d,\pp} \sim \phi\ , \quad D_d\sim \phi^{1+\alpha/2}\ .
\ee

The expansion of $H(\phi)$ around $\phi_\pp=0$ gives terms that are clearly zero except for a term proportional to
\be
\frac{(V-V_t)^{d/2}}{D_d^{d-1}}\left(\frac{V_\pp}{D_{d\pp}}+\frac{V_t}{V_t'}\right)\sim \phi^{1-\alpha(d-1)/2}\left[0 \times \frac{1}{\phi} + \phi +{\cal O}(\phi^{1+\alpha}) \right]\ ,
\label{ratio} 
\ee
that requires a more detailed analysis.
From Eqs.~(\ref{B}) and (\ref{a}), it follows that $0<\alpha<4/(d-1)$ and
so the quantity (\ref{ratio}) also goes to zero for $\phi\rightarrow 0$, ensuring that $H(0)=0$.

In the Minkowski case ($V_\pp=0$), we cannot simply take the limit $V_\pp\to 0$ in the previous AdS analysis. Solving the EOM for $V_t$, Eq.~(\ref{EoMVt}), in the small field regime, we find
\be
V_t(\phi) = -\frac12 m^2\phi^2\left[\frac{2}{W}+{\cal O}\left(\frac{1}{W^2}\right)\right]+{\cal O}(\phi^4)\ ,
\ee
where
\be
W\equiv W\left(\frac{2m}{d-1}\left(\frac{\phi}{\phi_c}\right)^{-2/(d-1)}\right)\ ,
\ee 
with $W(x)$ the product-log or Lambert $W$ function [solution of $W(x)e^{W(x)}=x$] and $\phi_c$ some constant field value (determined by the boundary conditions at $\phi_0$). At $x\to \infty$ we have 
\be
W(x)=\log x+\frac{(1-\log x)\log(\log x)}{\log x}+...
\ee
We then find that $V_t(\phi)\sim \phi^2/\log\phi$ at $\phi\to 0$. It follows that $V_t'\sim \phi/\log\phi$ and $D\sim -\phi/\log\phi$. Using these asymptotic behaviours we find that the non-trivial terms in (\ref{ratio}) go like
\be
\frac{(V-V_t)^{d/2}}{D_d^{d-1}}\left(\frac{V_t}{V_t'}\right)\sim \phi^2(-\log\phi)^{(d-2)/2}\to 0\ ,
\label{ratioM} 
\ee
and we find that $H(0)=0$ also in this case. In conclusion, for Minkowski and AdS decays we have proven that $S=\Delta S_E$.

\subsection{dS Vacua} 

Finally, consider the dS case, ($V_\pp>0$), for which $\xi_e$ is finite, with $\rho(0)=\rho(\xi_e)=0$. From the boundary conditions
on $\rho$, (\ref{BCdS}), we see that $S_{\rm GHY}=0$ (as it should be, given that the CdL instanton geometry is compact and has no boundary).
Then, the tunneling action reads 
\be
\Delta S_E=\int_{\phi_{0\pp}}^{\phi_0}{\it s}^{(d)}_E\,d\phi -S_{E\pp}\ .
\label{SEdS}
\ee
The action density for the CdL instanton part is obtained translating (\ref{SEs}) to field space and $V_t$ quantities, as before. One has
\be
{\it s}_E^{(d)}= -\frac{4\pi^{d/2}(d-1)^{d-1}}{(d-2)\Gamma(d/2)}\frac{V\left[2(V-V_t)\right]^{d/2-1}}{D_d^{d-1}}
\ .
\ee
The action for the background false vacuum is finite and can be calculated exactly simply plugging $\rho_\pp(\xi)=R_\pp \sin(\xi/R_\pp)$, with  $\xi_e=\pi R_\pp$ in (\ref{SEs}). One gets
\be
S_{E\pp}=-\frac{4\pi^{(d+1)/2}R_\pp^{d-2}}{\kappa\Gamma[(d-1)/2]}\ .
\label{SEp}
\ee

For this dS case, the relation between Euclidean and $V_t$ action
densities is
\be
{\it s}^{(d)} - {\it s}^{(d)}_E = \frac{dH_0}{d\phi}\ , 
\label{senewoldS}
\ee
with $H_0(\phi)$ given by $H(\phi)$ in (\ref{H}) without the $V_\pp$ term
\bea
H_0&=&
\frac{2(d-1)^d\left[2\pi(V-V_t)\right]^{d/2}}{(d-2)\Gamma(d/2+1) D_d^{d-1}}\left[\frac{V_t}{V_t'}\, _2F_1\left(1/2,1;d/2+1;1-D_d^2/V_t'{}^2\right)\right]\nonumber\\
&-&\frac{2\pi^{(d+1)/2}R_t^{d-2}}{\kappa \Gamma[(d-1)/2]}\left(1+\frac{V_t'}{|V_t'|}\right)\ .
\label{H0}
\eea
As for the previous cases, (\ref{senewoldS}) holds on-shell.

To prove the equivalence between the tunneling action for dS decay calculated in the Euclidean formalism, as given by (\ref{SEdS}), and the tunneling potential formalism, as given by (\ref{SdSalt}), we make use of (\ref{senewoldS}) integrated in the CdL interval $(\phi_{0\pp},\phi_0)$. We get
\be
 S - \Delta S_E  = H_0(\phi_0) - H_0(\phi_{0\pp}) 
 -\frac{4\pi^{(d+1)/2}R_{0\pp}^{d-2}}{\kappa\Gamma[(d-1)/2]}
\ .
\label{senewoldSint}
\ee
Now, at $\phi_0$ one still has $D_d=-V_t'$, so that $H_0(\phi_0)=0$,
while at $\phi_{0\pp}$ one has $V_t'>0$ so that $D_d=V_t'$, and \be
H_0(\phi_{0\pp})=\frac{4\pi^{(d+1)/2}R_{0\pp}^{d-2}}{\kappa\Gamma[(d-1)/2]}\ .
\ee 
Plugging this in (\ref{senewoldSint}) leads to the claimed equality $ S = \Delta S_E  $.

\section{More General Action \label{app:gen}} 
\noindent

In this appendix we obtain the tunneling action density in the $V_t$ formalism for a more general action of the form
\be
S_E=\int d^dx\sqrt{g}\left[G(\phi)R+\frac12Z(\phi)g^{\mu\nu}\partial_\mu\phi\partial_\nu\phi+V(\phi)\right]+S_{\rm GHY}\ ,
\label{SEG}
\ee
where we allow for a field-dependent non-canonical kinetic term parametrized by $Z(\phi)$  and a nonminimal coupling function $G(\phi)$.

It is straightforward to get rid of Euclidean quantities in terms of $V_t$ dependent ones and repeat the derivation of the main text. Skipping the details we present the main results.
The generalized $\rho(\xi)$ is
\be
\rho = (d-1) \frac{\sqrt{2(V-V_t)}}{D_d}\ ,
\ee
with
\be
D_d  \equiv \sqrt{\frac{\hat{V}_t'{}^2}{\hat Z}-2\frac{(d-1)}{(d-2)G} (V-V_t)V_t}\ ,
\ee
where $V_t$ has been defined by
\be
V_t\equiv V -\frac12 \hat Z \dot\phi^2\ ,
\ee
and
\be
\hat{V}_t'\equiv V_t'-\frac{d\, V_t G'}{(d-2)G}\ , \quad
\hat Z\equiv Z-\frac{2(d-1)G'{}^2}{(d-2)G}\ .
\ee

The EoM differential equation for $V_t$ reads:
\bea
0&=&
V_t'\left\{-d\, V_t'+(d-1)\left[V'+(V-V_t)\frac{\hat Z'}{\hat Z}\right]+(d+1)\left(2\frac{d-1}{d-2}V-V_t\right)\frac{G'}{G}\right\}
\nonumber\\
&+&\frac{d}{d-2}\frac{V_tG'}{G}\left\{-(d-1)V'+\left[(2d-1)V_t-2\frac{(d-1)^2}{d-2}V\right]\frac{G'}{G}
\right\}\nonumber\\
&-&2(d-1)(V-V_t)\left\{V_t''+\frac{\hat Z}{G}\left(V_t-\frac{d-1}{d-2}V\right)+\frac{d\, V_t}{d-2}\left[
\frac12\frac{\hat Z' G'}{\hat Z G}-\frac{G''}{G}\right]\right\}\ .
\label{VtEoMg}
\eea

This EoM can be rewritten in a simpler form in terms of $D_d$ as
\be
\frac{d}{d\phi}\log D_d = \frac{1}{2(V-V_t)}\left[V'-\frac{d\, V_t'}{d-1}  +\left(\frac{2V}{d-2}-\frac{V_t}{d-1}\right)\frac{G'}{G}\right]\ .
\ee

The action density is
\be
{\it s}^{(d)}=\frac{\sqrt{\pi}}{\Gamma[(d+1)/2]}\left[
\frac{(d-1)(d-2)\pi}{-|V_t|/ G}\right]^{d/2}(\hat V_t'+|\hat V_t'|)
 +\, _2F_1\left(\frac{d-1}{2},\frac{d}{2},\frac{d+2}{2},1-\frac{\hat Z D_d^2}{\hat V_t'{}^2}\right) {\it s}_0^{(d)}\, \ ,
\label{newSEgen}
\ee
with
\be
{\it s}_0^{(d)}=\frac{(d-1)^{(d-1)}\left[2\pi\hat Z (V-V_t)\right]^{d/2}}{\Gamma(1+d/2)|\hat V_t'|^{d-1}} \ ,
\ee
being the action density without gravity.
For the $d=4$ case one gets
\be
{\it s}^{(4)} =24\pi^2G^2\hat Z^{1/2}
\frac{(D_4+\hat V_t'/\hat Z^{1/2})^2}{V_t^2 D_4}\ .
\ee

The particular combinations of the functions $Z,V,G$ (and their derivatives) that appear in ${\it s}^{(d)}$ guarantee that the action density transforms in the right way under Weyl rescalings of the metric.
To see this explicitly, consider the Weyl rescaling to Einstein
frame
\be
g_{\mu\nu} = \left(\frac{G_E}{G}\right)^{\frac{2}{d-2}}g^E_{\mu\nu}\ ,
\ee
 in which $G_E=-1/(2\kappa)$.  One gets
 \be
 \hat Z_E = \frac{G_E}{G}\ \hat Z\ , \quad
 V_E =  \left(\frac{G_E}{G}\right)^{\frac{d}{d-2}} V\ ,
 \quad
 (\dot \phi^2)_E = \left(\frac{G_E}{G}\right)^{\frac{2}{d-2}}\dot \phi^2\ , 
 \ee
from which it follows
 \be
(V_t)_E =  \left(\frac{G_E}{G}\right)^{\frac{d}{d-2}} V_t\ ,
 \quad
 (\hat V'_t)_E =  \left(\frac{G_E}{G}\right)^{\frac{d}{d-2}} 
 \hat V_t'\ ,
 \quad
 D^2_{dE} =  \left(\frac{G_E}{G}\right)^{\frac{d+2}{d-2}} D_d^2\ .
 \ee
These transformation properties imply that the argument of the hypergeometric function and the tunneling action are invariant under the Weyl rescaling. This is needed to give physical meaning
to the decay rate, which should be frame-independent.

\section{Thin-Wall Action via Euclidean Formalism\label{app:thwE}}

In the Euclidean formalism, the thin-wall bounce for the decay of a  false vacuum is simply a ball
with $V\simeq V_\mm$ and $\phi_B\simeq \phi_\mm$ and radius $\rho_B=R_B$ (to be determined) surrounded by false vacuum with $V=V_\pp$ and $\phi_B=\phi_\pp$. The corresponding thin-wall action can be obtained using the general expression (\ref{SE}) to calculate the difference $\Delta S_E = S_E[\phi_B,\rho_B] -S_E[\phi_\pp,\rho_\pp]$. It is convenient to discuss separately the different types of false vacua.

\subsection{Minkowski or AdS Vacua}

Outside the wall at $R_B$, the bounce solution coincides with the background and the contribution to $\Delta S_E$ is zero.
For the contribution to $\Delta S_E$ from inside the wall we have
$\rho_B(\xi)=R_\mm \sinh(\xi/R_\mm)$ and $\rho_\pp(\xi)=R_\pp \sinh((\xi+c)/R_\pp)$, where $c$ is a constant shift to match both solutions at the wall. This leads to
\be
\Delta S_{E,in}=-\frac{(d-1)(d-2)V_{S,d-1}}{\kappa} \int_0^{R_B}\rho^{d-3}\left[\sqrt{1+\rho^2/R_\mm^2}\, d\rho- \sqrt{1+\rho^2/R_\pp^2}\right]\, d\rho
\ .
\ee
In addition, there is a contribution from the wall, as both $\dot\phi^2$ and $V$ have a delta function at $\rho=R_B$. The wall tension is
\be
\sigma\equiv \int_{\xi_B-\delta}^{\xi_B+\delta} \left[\frac12 \dot\phi_B^2+V(\phi_B)-V_\pp\right]\, d\xi\ ,
\ee
where $\xi_B$ is such that $\rho_B(\xi_B)=R_B$. To see that this definition agrees with the one in the $V_t$ formalism, note that, in the thin-wall limit, $V-\dot\phi^2/2\simeq V_\pp\simeq V_\mm$ is approximately constant, so that 
\be
\sigma= \int_{\xi_B-\delta}^{\xi_B+\delta} \dot\phi_B^2\, d\xi
=\int_{\phi_\mm}^{\phi_\pp} \dot\phi_B^2\, \frac{d\phi}{\dot\phi_B}=\int_{\phi_\pp}^{\phi_\mm} \sqrt{2(V-V_t)}\, d\phi\ ,
\ee
reproducing (\ref{sigmaVt}).
The contribution to the Euclidean action from the wall is then
\be
S_{E,wall}=V_{S,d-1}\sigma R_B^{d-1}\ .
\ee

The final result for the tunneling action is therefore
\be
\Delta S_E=V_{S,d-1}\left\{\sigma R_B^{d-1}-\frac{(d-1)(d-2)}{\kappa}
\int_0^{R_B}\rho^{d-3}\left[\sqrt{1+\rho^2/R_\mm^2}\, - \sqrt{1+\rho^2/R_\pp^2}\right]\, d\rho
\right\}\ .
\ee
The value of the bounce radius can be obtained from extremizing this action, $d\Delta S_E/dR_B=0$, which gives 
\be
R_B=\frac{(d-2)}{\kappa\sigma}\left[\sqrt{1+R_B^2/R_\mm^2}\, - \sqrt{1+R_B^2/R_\pp^2}\right]\ ,
\ee
and this is  in agreement with $R_B$ calculated in $V_t$ formalism, see (\ref{RB}).\footnote{The same result follows from integrating $\ddot\rho=\kappa\rho[(2-d)\dot\phi^2-2V]/[(d-1)(d-2)]$ across the wall.\label{dotrhojump}} Having determined $R_B$, we finally perform the integrals appearing in the action and get
\be
\Delta S_E=V_{S,d-1}\left\{
\sigma R_B^{d-1} -\frac{(d-1)}{\kappa}R_B^{d-2}\left.\left[
{}_2F_1(-1/2,d/2-1,d/2,-R_B^2/R^2)
\right]\right|_{R_\pp}^{R_\mm}
\right\}\ .
\ee
To show that this agrees with the result (\ref{SthwR}) of the $V_t$ formulation one simply needs to use the hypergeometric function identity
\be
{}_2F_1(-1/2,d/2-1,d/2,z)=\sqrt{1-z}+\frac{z}{d}\,{}_2F_1(1/2,d/2,d/2+1,z)\ ,
\ee
and the relation between $\sigma$ and $R_B$.

\subsection{dS Vacua}
As explained in the main text, in the thin-wall regime tunneling proceeds from $\phi_{0\pp}\simeq \phi_\pp$ to $\phi_0\simeq \phi_\mm$. The metric function $\rho$ is not monotonic but is rather a concave function with two zeros and, in what follows, one should pay attention to the change of sign of $\dot\rho$. We start discussing dS to dS transitions and will complete the discussion with  decays from dS to Minkowski or AdS at the end.  For the $d=4$ case, see {\it e.g.} \cite{Weinberg,EFT}.

\begin{figure}[t!]
\begin{center}
\includegraphics[width=0.6\textwidth]{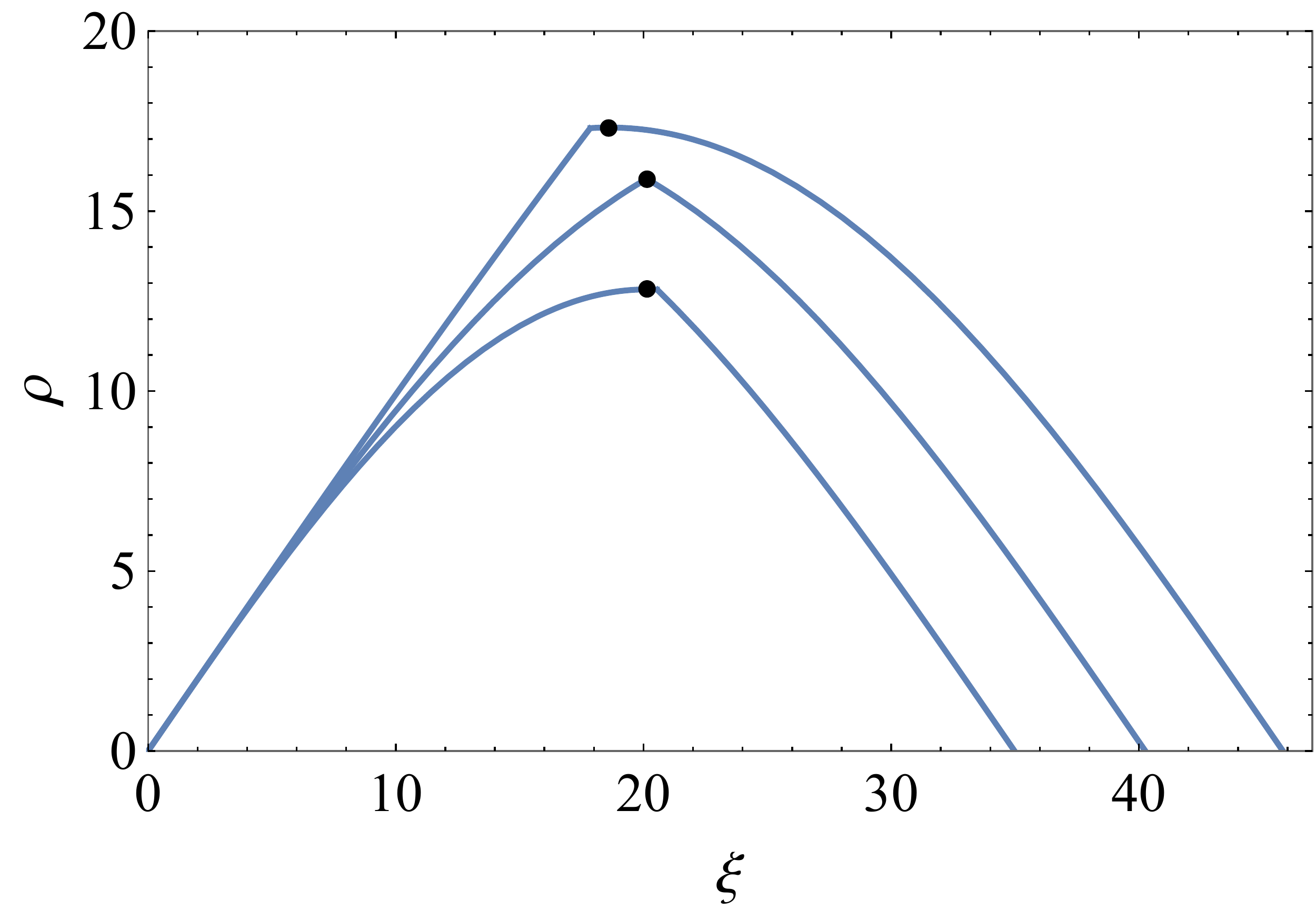}
\end{center}
\caption{Metric function $\rho(\xi)$ for the three cases of dS decay: case (1) with $V_\mm<V_\pp-\delta V$  (upper curve); case (2) with $V_\mm>V_\pp+\delta V$ (lower curve) and  case (3) with $V_\mm=V_\pp$ (central curve). The location of the maximum of each curve is marked by a black dot. These curves correspond to the same examples of Figure \ref{fig:dSCasesVt}.
\label{fig:dSCasesrho}
}
\end{figure}

The bounce field configuration has $\phi_B(\xi<\xi_B)\simeq \phi_\mm$
and $\phi_B(\xi>\xi_B)\simeq \phi_\pp$ with a rapid transition between the two values at the wall, located at $\xi_B$. The metric function is $\rho_B(\xi<\xi_B)=R_\mm \sin(\xi/R_\mm)$ and $\rho_B(\xi>\xi_B)=R_\pp \sin((\xi+c)/R_\pp)$,
and $c$ is a constant fixed by the continuity condition $\rho_B(\xi_B+\delta)=\rho(\xi_B-\delta)=R_B<R_\pm$ for $\delta\ll\xi_B$.  The jump in $\dot\rho$ at $\xi_B$ can be obtained integrating $\ddot\rho$ across the wall, see footnote \ref{dotrhojump}. One gets
\be
\dot\rho(\xi_B+\delta)-\dot\rho(\xi_B-\delta)=-\frac{\kappa\sigma R_B}{(d-2)}\ ,
\ee
showing that the slope of $\rho$ drops at the wall. This negative drop forces only three cases for the qualitative shape of $\rho(\xi)$
depending on whether its maximum occurs inside the bubble, outside it or at the wall. Figure~\ref{fig:dSCasesrho} shows the $\rho$ profiles in the three cases: (1) For $V_\mm<V_\pp-\delta V$, 
where $\delta V=\kappa_d\sigma^2/2$,
one has $\xi_B<\pi/2$  with the result that $\rho$ peaks at ${\rm max}(\rho)=R_\pp$ for $\xi>\xi_B$, when $\phi\simeq \phi_\pp$. In this case, $\dot\rho(\xi_B\pm\delta)>0$. (2) For 
$V\mm>V_\pp+\delta V$, one has $\xi_B>\pi/2$ with $\rho$ peaking at ${\rm max}(\rho)=R_\mm$ for $\xi<\xi_B$, when $\phi\simeq \phi_\mm$. In this case, $\dot\rho(\xi_B\pm\delta)<0$. (3)
For $|V_\pp- V_\mm|<\delta V$, $\rho$ peaks (with a cusp) at $\xi_B$, right at the wall. In this case $\dot\rho(\xi_B-\delta)>0$ and $\dot\rho(\xi_B+\delta)<0$. These three cases exactly correspond to those discussed in the text for the $V_t$ approach. Note in particular that the maximum of $\rho$ occurs at the same field value as the maximum of $V_t$, see (\ref{drho}).

From the previous discussion we get, for cases (1) and (2)
\be
\sqrt{1-R_B^2/R_{\rm max}^2}-\sqrt{1-R_B^2/R_{\rm min}^2}=\frac{\kappa\sigma R_B}{(d-2)}\ ,
\ee
where $R_{\rm max}\equiv {\rm max}(R_\pp,R_\mm)$ and
 $R_{\rm min}\equiv {\rm min}(R_\pp,R_\mm)$. While, for case (3),
\be
\sqrt{1-R_B^2/R_\pp^2}+\sqrt{1-R_B^2/R_\mm^2}=\frac{\kappa\sigma R_B}{(d-2)}\ .
\ee
These formulas are in agreement with (\ref{sigmadS}).

It is then straightforward to calculate the Euclidean action for the bounce and the background in all three cases and massage it conveniently into this single formula valid for all cases: 
\bea
\Delta S_E &=&
\left\{\frac{(\Delta V-\delta V)R_B R_\pp^{d-2}}{\kappa\sigma\sqrt{1-R_B^2/R_\pp^2}}\left[\left.  x^{d/2-1}{}_2F_1(-1/2,d/2-1,d/2,x)\right|_{x=1}^{x=R_B^2/R_\pp^2}\right]\right.
\nonumber\\
&&-
\frac{(\Delta V+\delta V)R_B R_\mm^{d-2}}{\kappa\sigma\sqrt{1-R_B^2/R_\mm^2}}\left[\left.  x^{d/2-1}{}_2F_1(-1/2,d/2-1,d/2,x)\right|_{x=1}^{x=R_B^2/R_\mm^2}\right]\nonumber\\
&&+\left.\frac{\sqrt{\pi}\Gamma(d/2)}{\kappa \Gamma((d-1)/2)}\left(R_\pp^{d-2}-R_\mm^{d-2}\right)+\sigma R_B^{d-1}\right\}  V_{S,d-1}\ .
\eea
This agrees with the $V_t$ result as can be checked using hypergeometric function identities and the relation for $\sigma$.

The case of a dS vacuum decaying to Minkowski or AdS is an straightforward generalization of the previous result to $R_\mm\to \infty$ and $R_\mm^2\to R_\mm^2<0$, respectively, as in the discussion at the end of subsection \ref{sec:dSthw}.


\end{document}